\documentclass[%
 reprint,
superscriptaddress,
 amsmath,
 amssymb,
 aps,
 prx,
 showpacs, 
 floatfix,
]{revtex4-2}

\usepackage{amsmath}
\usepackage{graphicx}
\usepackage{dcolumn}
\usepackage{bm}
\usepackage{orcidlink}

\begin{document}

\preprint{APS/123-QED}

\title{Path-integral approach to quantum acoustics}

\author{Joost V. de Nijs \orcidlink{0009-0005-4827-2106}}

 \affiliation{Department of Chemistry and Chemical Biology, Harvard University, Cambridge,
Massachusetts 02138, USA}
\affiliation{%
    Faculty of Applied Sciences, Delft University of Technology, 2628 CD Delft, Netherlands
}%

\author{Anton M.~Graf \orcidlink{0000-0002-0573-8907}}
\affiliation{Harvard John A. Paulson School of Engineering and Applied Sciences,
Harvard University, Cambridge, Massachusetts 02138, USA}
\affiliation{Department of Chemistry and Chemical Biology, Harvard University, Cambridge,
Massachusetts 02138, USA}
\affiliation{Department of Physics, Harvard University, Cambridge, Massachusetts 02138, USA}

\author{Eric J.~Heller \orcidlink{0000-0002-5398-0861}} 
\affiliation{Department of Chemistry and Chemical Biology, Harvard University, Cambridge,
Massachusetts 02138, USA}
\affiliation{Department of Physics, Harvard University, Cambridge, Massachusetts 02138, USA}

\author{Joonas~Keski-Rahkonen \orcidlink{0000-0002-7906-4407}}
\affiliation{Department of Chemistry and Chemical Biology, Harvard University, Cambridge,
Massachusetts 02138, USA}
\affiliation{Department of Physics, Harvard University, Cambridge, Massachusetts 02138, USA}

\date{\today}

\begin{abstract}
\noindent
A path-integral approach to quantum acoustics is developed here. In contrast to the commonly utilized particle perspective, this emerging field brings forth a long-neglected yet essential wave paradigm for lattice vibrations. Within the coherent state picture, we formulate a non-Markovian stochastic master equation that captures the exact dynamics of any system with coupling linear in the bath coordinates and nonlinear in the system coordinates.
We further demonstrate the capability of the presented master equation by applying the corresponding procedure to the well-known Fr{\"o}hlich model. In general, we establish a solid foundation for quantum acoustics as a kindred framework to quantum optics, while paving the way for deeper first-principles explorations of nonperturbative system dynamics driven by lattice vibrations.
\end{abstract}

\maketitle


\section{Introduction}
\noindent
A wide range of physical properties of condensed matter involve, in some way, electrons interacting with the crystal lattice, notably as in the process that determines electrical resistance. Since the advent of second quantization, lattice vibrations have been described as a quantum field in the Fock state formalism~\cite{Mahan_book, Giustino_rev.mod.phys_89_015003_2018}, essentially making the particle (phonon) aspect of lattice vibrations the primary focus. Apart from an early stage when lattice vibrations were initially treated as a classical field, the predominant avenue to investigate the electron-lattice dynamics has relied on perturbation theory within the number state picture (see, e.g., Refs.~\cite{Mahan_book, Giustino_rev.mod.phys_89_015003_2018, ashcroft1976solid, kittel2018introduction}). For instance, the conventional Bloch-Gr{\"u}neisen theory~\cite{Bloch_z.phys_59_208_1930, Gruneisen_ann.phys_16_530_1933} successfully describes the temperature dependence of electrical resistivity of most metals by
taking the scattering of electrons from acoustic phonons into account: the interaction of an electron with lattice vibrations involves phonon creation or annihilation within the first-order perturbation theory, whereas the higher-order interactions are approximated as an incoherent and uncorrelated chain of first-order events employing the Boltzmann transport theory.

\begin{figure}[t!]
\centering
\includegraphics[width=\linewidth]{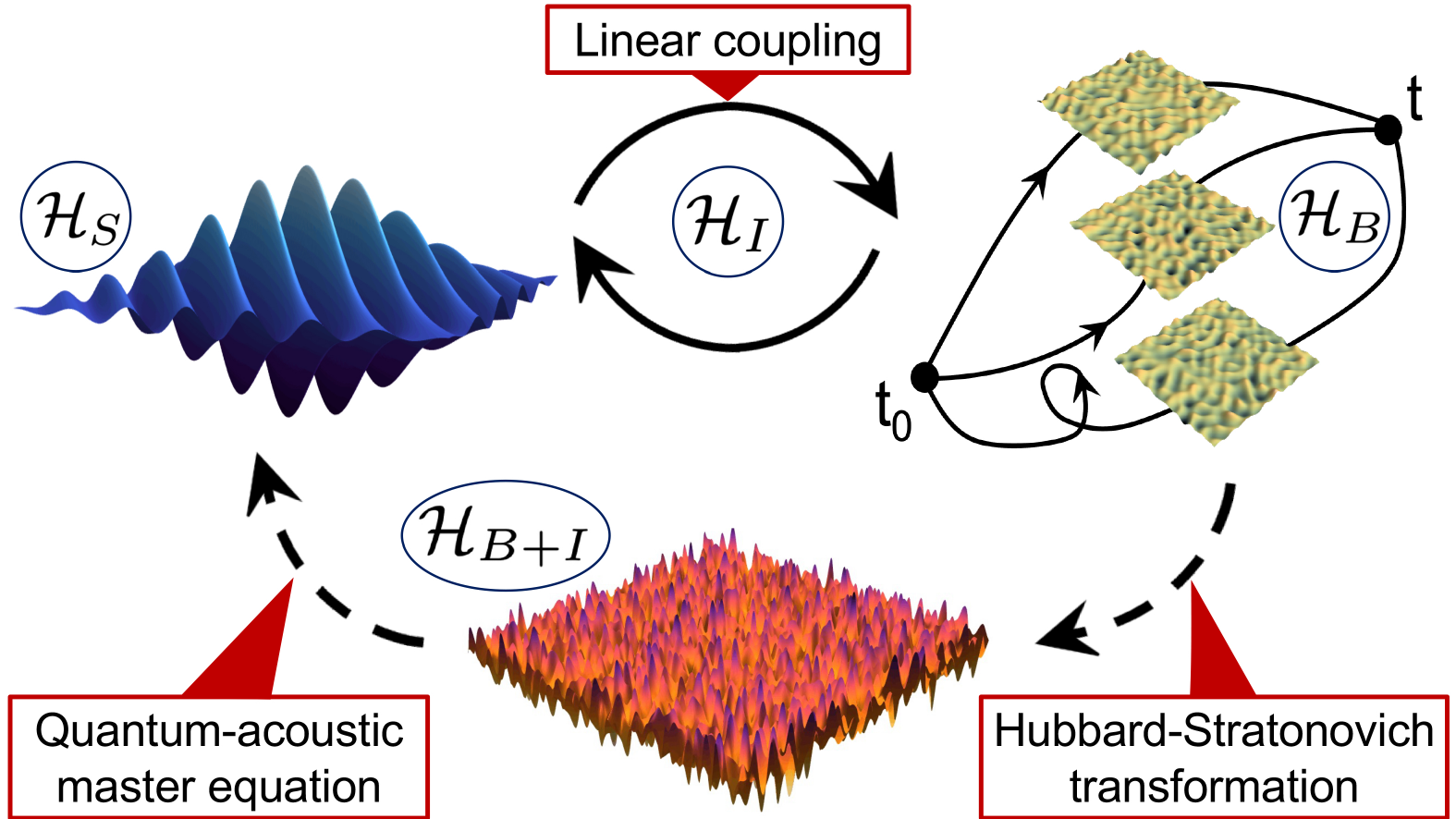}
\caption{The figure summarizes the general scheme behind the path-integral approach for the quantum-acoustical master equation. We examine a general system $\mathcal{H}_S$, depicted in blue, interacting with a harmonic lattice bath $\mathcal{H}_B$, under an arbitrary linear coupling $\mathcal{H}_I$. The time evolution of the system can be expressed with a path integral over all possible bath states, or equivalently all possible mean-field (deformation) potentials, as shown in the top right, which further results in a stochastic pseudo-potential $\mathcal{H}_{B+I}$, shown in red.
}
\label{fig:promo}
\end{figure}

In contrast, the coherent state representation of the lattice vibrations replaces the burden of keeping track of individual phonons with information about the
phase and amplitude of each vibrational normal mode. This program follows the quantum optics pathway pioneered by Glauber~\cite{gauber_phys_rev_131_2766_1963}, establishing a long-neglected but essential wave perspective for lattice vibrations, called \emph{quantum acoustics}. In other words, there exists a parallel world of lattice vibrations akin to quantum optics~\cite{walls2007quantum, scully1997quantum}, with the phonon playing the role of the photon, and acoustic waves playing the role of electromagnetic waves.  

Although the coherent state picture for lattice vibrations developed in Ref.~\cite{Donghwan_phys.rev.b_106_054311_2022} is the dual counterpart to the traditional number state description, it provides a different premise for electron-lattice dynamics, with its own set of possible approximations. For example, the quasi-classical limit of quantum acoustics unveils a real-space, time-dependent description of electron-lattice interaction in terms of a deformation potential (see Fig.~\ref{fig:promo}). A very similar notion was introduced in 1957 by Hanbury Brown and Twiss for the vector potential of a blackbody field~\cite{HanburyBrown_proc.r.soc.a_242_200_1957}, with the essential difference of missing the ultraviolet cutoff, i.e., the Debye wavenumber in the definition of our deformation potential originating from the minimal lattice spacing. On the other hand, in the 1950s, Bardeen and Shockley considered dynamical lattice distortions in nonpolar semiconductors, but were constrained to a perturbative proxy frozen in time~\cite{Shockley_phys.rev_77_407_1950, Bardeen_phys.rev_80_72_1950}.  However, this pathway using lattice vibrations languished, with attention centered instead on perturbation theory methods, commonly rooted in the number state formalism. 

Over 70 years later, the quantum-acoustical framework implied by Bardeen and Shockley has been revived, demonstrating its ability to recover the conventional results of the Bloch-Grüneisen theory~\cite{Bloch_z.phys_59_208_1930, Gruneisen_ann.phys_16_530_1933} for normal metals in the perturbation theory limit, but from the \emph{time-dependent} vantage point. This aspect is particularly of importance for the deformation potential approach where the lattice deformation is intrinsically dynamic. 

The quantum-acoustic paradigm has already shed light on the dynamical birth and life of acoustic polarons~\cite{aydin_pnas_122_e2426518122_2025} and low-energy tails of spectral densities~\cite{Donghwan_phys.rev.B_107_224311_2023}, as well as addressed several puzzles associated with  so-called strange metals, including the violation of the Mott-Ioffe-Regel limit~\cite{aydin_pnas_121_e2404853121_2024}, the elusive appearance of a displaced Drude peak~\cite{KeskiRahkonen_phys.rev.lett_132_186303_2024}, and the mystery of linear-in-temperature resistivity across a wide temperature range, all tied to the enigmatic yet pervasive Planckian bound~\cite{aydin_pnas_121_e2404853121_2024}.

Despite its relatively recent introduction in comparison with the number-state alternative, the program of quantum acoustics in its original formulation~\cite{Donghwan_phys.rev.b_106_054311_2022} is rapidly gaining traction in solid-state physics, as overviewed above. Here, we further substantiate this fascinating direction and point the way forward to improvements. We focus on understanding and modeling of coherent lattice vibrations from first principles by emphasizing the long-missing coherent state scheme~\cite{Donghwan_phys.rev.b_106_054311_2022}. Following the vision of Feynman~\cite{feynman1965quantum}, we present the path-integral formulation of quantum acoustics, as illustrated in Figure~\ref{fig:promo}. Specifically, we introduce a stochastic master equation, akin to its quantum-optical counterparts, which encompasses the \emph{exact} dynamics for a large class of electron systems interacting with a harmonic lattice under a general coupling. 

Over the years, a manifold of (numerically) exact methods for open quantum dynamics has emerged (see, e.g., Refs.~\cite{StockburgerColloquium, de.vega_rev.mod.phys_89_015001_2017, Breuer_Rev.Mod.Phys_88_021002_2016, Weiss_book}). For instance, a key advantage of utilizing the path-integral formalism~\cite{feynman1965quantum} is that the environmental degrees of freedom can be integrated out exactly, while operator-ordering issues do not arise explicitly~\cite{StockburgerColloquium, Weiss_book}. The starting point is the path-integral representation, enabling a nonperturbative treatment of Gaussian quantum dissipation, as pioneered by Feynman and Vernon~\cite{Feynman_ann.phys_24_118_1963}, and developed extensively thereafter (see, e.g., Refs.~\cite{Weiss_book, Legget_Rev.Mod.Phys_59_1_1987, Grabert_Phys.Rep_168_115_1988}, thereby substantiating the path-integral approach as a powerful tool for studying~\cite{schulman_book, Weiss_book, Wong_Commun.Comput.Phys_15_853_2014}, e.g., solid-state and molecular systems (see, e.g., Refs.~\cite{Althorpe_Annu.Rev.Phys.Chem_75_397_2024, Markland_Nat.Rev.Chem_2_0109_2018, Ananth_Annu.Rev.Phys.Chem_73_299_2022, Herrero_j.condens.matter.phys_26_233201_2014}).

Within the path-integral framework, the effect of the environment is fully encoded in the Feynman-Vernon influence phase, which has, for example, led to path-integral Monte Carlo methods for numerical simulations, such as showcased in Refs.~\cite{BarkerPathIntegralMC, Dornheim_Phys.Rep_744_1_2018, Ceperley_Rev.Mod.Phys_67_279_1995}. Their applicability is, nevertheless, limited by the infamous sign problem~\cite{Egger_DynamicalSignProblem, feynman1965quantum, Ceperley_J.Stat.Phys_63_1237_1991, Feynman_Int.J.Theor.Phys_21_467_1982, Fosdick_Phys.Rev_143_58_1966}, motivating alternatives to the approach that either retain the path-integral formulation while alleviating the sign problem~\cite{kundu_annu.rev.phys.chem_73_349_2022,foulkes_rev.mod.phys_73_33_2001}, or instead divert in the direction of master-equation techniques~\cite{tanimura_j.phys.soc.jpn_58_101_1989,StockburgerColloquium}. An example of the latter is provided by stochastic methods pioneered by Stockburger~\cite{Stockburger_vehm.phys_268_249_2001}, in which the influence functional is \emph{unraveled} through the introduction of auxiliary stochastic fields, yielding a numerically exact master equation. Conceptually, this procedure relies on Gaussian identities~\cite{Altland_book, Sieberer_Rep.Prog.Phys_79_096001_2016} that recast the time-nonlocal path-integral expression into an equivalent time-local equation of motion in an extended Liouville space~\cite{StockburgerColloquium}. Here, we adopt this strategy.

In a sense, we present an extension to Stockburger's method that subsequently lays a mathematically rigorous foundation for the emerging field of quantum acoustics. Previously, a similar scheme has been utilized to investigate spin-boson dynamics~\cite{stockburger_europhys.lett_115_40010_2016, stockburger_chem.phys.296.159.2004, stockburger_phys.rev.lett_88_170407_2002}, and to assess the accuracy of second-order perturbation theory for polarons~\cite{lee_j.chem.phys_136_204120_2012}. However, to the best of our knowledge, the case of non-bilinear coupling with discrete or coherent baths has not been explored, nor has this possibility been applied to a concrete physical problem. In addition to formulating the quantum-acoustical master equation, we further apply this strategy to the specific case of the venerable Fr{\"o}hlich model~\cite{Frochlich_adv.phys_3_325_1954, Frochlich_proc.r.soc.a_160_230_1937}, comparing it against the mean-field and perturbation theory approaches.

\section{Quantum-acoustical master equation}

We begin by considering a composite system of $\mathcal{H}=\mathcal{H}_S+\mathcal{H}_B+\mathcal{H}_I$. It entails a generic quantum system of charge carriers $\mathcal{H}_S$ that interacts with a lattice $\mathcal{H}_B$ modeled as a bath of harmonic oscillators. By denoting the coordinates of a normal mode as $\boldsymbol{X}_{\boldsymbol{q}} =\left[\begin{matrix} x_{\boldsymbol{q}} & p_{\boldsymbol{q}}  \end{matrix}\right]$, the lattice bath $\mathcal{H}_B$ and the linear coupling $\mathcal{H}_I$ are defined as $\mathcal{H}_B = \frac{1}{2}\sum_{\boldsymbol{q}} \omega_{\boldsymbol{q}}\Big (\boldsymbol{X}_{\boldsymbol{q}}\cdot\boldsymbol{X}_{\boldsymbol{q}}^T \Big)$ and $\mathcal{H}_I = \frac{1}{2}\sum_{\boldsymbol{q}} \Big( \boldsymbol{g}_{\boldsymbol{q}}(\boldsymbol{r}) \cdot \boldsymbol{X}_{\boldsymbol{q}}^T \Big)$
where $\bm g_{\bm q}(\bm r) = \left[\begin{matrix} g^0_{\bm q}(\bm r) & g^1_{\bm q}(\bm r) \end{matrix}\right]$ consists of model-specific functions depending only on the system coordinates $\boldsymbol{r}(t)$. 

In a fashion similar to quantum optics~\cite{walls2007quantum, scully1997quantum}, we derive the corresponding quantum master equation that governs the exact dynamics of the reduced density matrix $\rho_S$ of a system of interest, starting from the path-integral formulation. In the following, we only outline the key definitions and steps, while the more complete derivation can be found in Appx~\ref{Appendix:master_equation}.

We start by assuming that the initial density matrix $\rho$ of the composite system is separable, that is, $\rho(t=0)=\rho_S\otimes \rho_B$, which is a common starting point~\cite{Weiss_book}. However, the composite system is not doomed to stay in this product form throughout its evolution, contrary to mean-field approaches, such as those presented in Refs.~\cite{Donghwan_phys.rev.b_106_054311_2022, aydin_pnas_121_e2404853121_2024,aydin_pnas_122_e2426518122_2025}.

Employing the independence of normal modes, the total set of lattice vibrations can be described as a product state of the coherent states of the normal modes, i.e., as a multimode coherent state~\cite{heller_J.Phys.Chem.A_123_4379_2019}. Following this strategy, we will model the lattice bath as multimode coherent state, resulting in the Gaussian density matrix $\rho_B( \boldsymbol{X},\boldsymbol{X}';\boldsymbol{X}^0)$. We however stress that while we choose to focus on the multimode coherent bath state $\rho_B( \boldsymbol{X},\boldsymbol{X}';\boldsymbol{X}^0)$, our procedures remain valid for a thermal bath as well, as described in Appx.~\ref{Appendix:master_equation} and in the earlier work of Ref.~\cite{Stockburger_vehm.phys_268_249_2001}. In short, a thermal initial state can be formulated as an ensemble average of multimode coherent states, meaning that it is given by the same ensemble average of the final states.

Given the initial state $\rho_S(t=0)$, the state of the system $\rho_S(t)$ at a later time $t$ is determined~\cite{feynman1965quantum, Weiss_book} by the propagator $J$ that is expressible in terms of a path-integral where the effect of the bath is incorporated in the influence functional $\mathcal{F}$~\cite{Feynman_ann.phys_24_118_1963}. However, it results in an action that is nonlocal in time. As a consequence, a direct numerical evaluation of the path integral, e.g., via path-integral Monte Carlo methods, encounters the dynamical sign problem~\cite{Egger_DynamicalSignProblem}. To circumvent this issue, we take advantage of the Hubbard–Stratonovich transformation~\cite{Hubbard_phys.rev.lett_3_77_1959,stratonovich_book}, which introduces auxiliary stochastic variables, thereby rendering the action local in time. This reformulation enables us to derive the following stochastic master equation for the system's density matrix $\tilde{\rho}_S$ in the form of
\begin{equation}
\label{eq:LvN}
\begin{split}
    i\frac{\partial}{\partial t}\tilde{\rho}_S&= \Big[\mathcal{H}_{S}+\mathcal{H}_{\text{mf}},\tilde\rho_S \Big]\\ &- \sum_{\boldsymbol{q}} \left( [\bm\eta_{\boldsymbol{q}}(t)\cdot \bm g_{\boldsymbol{q}},\tilde \rho_S]-\frac{1}{2}\{\bm\nu_{\boldsymbol{q}}(t)\cdot \bm g_{\boldsymbol{q}},\tilde \rho_S\} \right)
\end{split}
\end{equation}
where $\mathcal{H}_{\textrm{mf}}$ is the mean-field interaction Hamiltonian, and $\bm \eta _{\boldsymbol{q}}$ and $ \bm\nu_{\boldsymbol{q}}$ are noise variables with mean zero and covariance:
\begin{equation} \label{eq:noise_covariance_original}
\begin{split}
\Big \langle \bm \eta^T_{\bm q}(s) \cdot \bm \eta_{\boldsymbol{r}}(t')\Big \rangle_W &=\frac{1}{2} \bar{\boldsymbol{L}}_{\bm q}(t'-s)\delta_{\bm q,\boldsymbol{r}};
\\
\Big \langle \bm\eta^T_{\boldsymbol{q}}(s)\bm \cdot \bm \nu_{\bm r}(t')\Big \rangle_W &=i\bar{\boldsymbol{M}}_{\bm q}(t'-s)\Theta(s-t')\delta_{\bm q,\boldsymbol{r}};
\\
\Big \langle \bm\nu^T_{\bm q}(s) \cdot \bm\nu_{\boldsymbol{r}}(t')\Big \rangle_W &=0,
\end{split}
\end{equation}
where we have defined 
\begin{align}
\begin{split}
    \bar{\boldsymbol{L}}_{\bm q}(t'-s)&=
    \begin{bmatrix}
    \cos[\omega_{\bm q}(t'-s)] &-\sin[\omega_{\bm q}(t'-s)] \\\sin[\omega_{\bm q}(t'-s)] &\cos[\omega_{\bm q}(t'-s)] 
    \end{bmatrix},\\
    \bar{\boldsymbol{M}}_{\bm q}(t'-s)&=
    \begin{bmatrix}\sin[\omega_{\bm q}(t'-s)] &0 \\0 &-\sin[\omega_{\bm q}(t'-s)] \text{\hspace{0.05cm}}
    \end{bmatrix}.
\end{split}
\end{align}
 This yields the physical density matrix $\rho_S$ when averaged over the noise $\bm \eta _{\boldsymbol{q}}$ and $\bm\nu_{\boldsymbol{q}}$, whose collective distribution we call $W$, i.e, $\rho_S=\left<\tilde\rho_S\right>_W $. Notably, the derived master equation is valid beyond the weak system-bath coupling and the Markovian approximation. 

Furthermore, unlike Lindbladian alternatives~\cite{Weiss_book}, we can further decompose the master equation into two independent Schrödinger equations with non-Hermitian Hamiltonians. By applying the ansatz of $\tilde{\rho}_S(t=0)=|\psi_{+}\rangle\langle\psi_{-}|$, we can compute the evolution of these states $|\psi_{-}\rangle$ and $|\psi_{+}\rangle$ as
\begin{equation}
\begin{split}
    i \frac{\partial}{\partial t}|\psi_{\pm}\rangle =\Bigg(\mathcal{H}_{S}+\mathcal{H}_{B+I}^{\pm}\Bigg)|\psi_{\pm}\rangle.
\end{split}   
\label{eq:time evolution}
\end{equation}
where we have defined a stochastic pseudopotential accommodating the lattice interaction on the system in the form of
\begin{equation}
\begin{split}
    \mathcal{H}_{B+I}^{\pm} = \left[\mathcal{H}_{\textrm{mf}} - \sum_{\bm q} \Big(\bm \eta_{\bm q}(t)\cdot \bm g_{\bm q} \mp \frac{1}{2}\bm \nu_{\bm q}(t)\bm \cdot \bm g_{\bm q} \Big)\right]^{-/\dagger}. 
\end{split}   
\end{equation}
Succinctly, to employ the described quantum-acoustical master equation approach, one first selects the system Hamiltonian $\mathcal{H}_S$ and sets up the form of the interaction, i.e., the functions $\bm g_{\bm q}$. Then, the system state $|\psi_{\pm}\rangle$ and coherent state parameters $\bm X^0_{\bm q}$ are initialized. The evolution of the system  $|\psi_{\pm}(t)\rangle$ is determined according to Eq.~\ref{eq:time evolution}. The expectation value of any system observable $\hat O(t)$ can consequently be evaluated as 
\begin{equation}\label{eq:expectation_value}
    \langle\hat O \rangle (t) =\left<\langle\psi_-(t)|\hat O|\psi_+(t) \rangle\right>_W.
\end{equation}

Interestingly, even though we have traced out the bath and thus become agnostic to its time evolution, we are still able to recover the expected values of certain observables via the Ehrenfest theorem~\cite{ehrenfest_z.phys_45_455_1927}. For instance, the expected values of the position and momentum operators for a lattice mode $\boldsymbol{q}$ are given by
\begin{align}
\label{eq: Ehrenfest}
    \frac{d}{dt} \boldsymbol{X}_{\boldsymbol{q}}(t) = \sigma_2\cdot\left(\omega_{\boldsymbol{q}}  \boldsymbol{X}_{\boldsymbol{q}}(t)  +\frac{1}{2}\langle \boldsymbol{g}_{\bm q} \rangle(t) \right),
\end{align}
where $\sigma_2$ is the Pauli matrix and the expectation values of $\bm g_{\bm q}(\bm r)$ are determined by Eq.~\ref{eq:expectation_value}. Therefore the utilization of the Ehrenfest theorem leads to coupled linear differential equations mimicking the classical motion of a bath composed of harmonic oscillators.

\section{Simple quantum-acoustical model}

We further showcase the applicability of the formalism developed above within the context of the standard Fr{\"o}hlich model~\cite{Frochlich_adv.phys_3_325_1954, Frochlich_proc.r.soc.a_160_230_1937}. A full description of the simulations can be found in Appx.~\ref{Appendix:simulation_details}. The mean-field component of the model corresponds to the deformation potential $V_{\textrm{def}}$, previously studied in Refs.~\cite{Donghwan_phys.rev.b_106_054311_2022, KeskiRahkonen_phys.rev.lett_132_186303_2024, aydin_pnas_121_e2404853121_2024, aydin_pnas_122_e2426518122_2025, zimmermann_entropy_26_552_2024}. Since the deformation potential $V_{\textrm{def}}$ averages to zero, it is best characterized by its
root-mean-square $\Delta V_\textrm{def}$. Comparing this scale with the Fermi energy $E_F$ yields the classification $\bar{K} = E_F/\Delta V_\textrm{def}$ for the electron–lattice coupling strength.

For example, normal metals fall within the perturbative regime ($\bar{K} \gg 1$), while the class of compounds known as strange or bad metals exemplifies the nonperturbative electron-lattice dynamics ($\bar{K} \lesssim 1$). In this work, we focus on copper and the prototypical strange metal Bi2212. Computational details are provided in Appx.~\ref{Appendix:simulation_details}, based on experimental data and consistent with previous studies~\cite{Donghwan_phys.rev.b_106_054311_2022, KeskiRahkonen_phys.rev.lett_132_186303_2024, aydin_pnas_121_e2404853121_2024, aydin_pnas_122_e2426518122_2025, zimmermann_entropy_26_552_2024}. However, we want to stress that the physics we find below transcends the material-specific constraints.

\begin{figure}[h!]
    \includegraphics[]{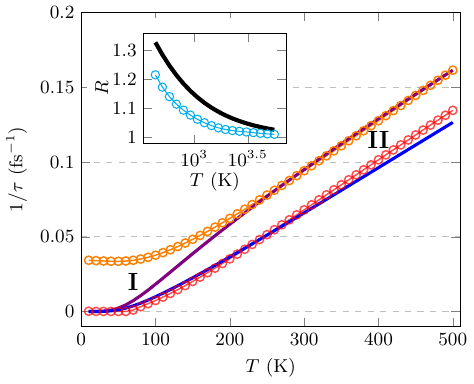}
    \caption{Inverse relaxation time as a function of temperature in the weak coupling limit, computed by employing the stochastic method $\tau_{\textrm{st}}$ (orange dots) and the mean-field approach $\tau_{\textrm{mf}}$ (red dots). These results are compared with the corresponding predictions $\tilde{\tau}_{\textrm{mf}}$ and $\tilde{\tau}_{\textrm{st}}$ from perturbation theory (purple and blue lines). In addition, the inset shows the relative values $R$ of the stochastic and mean-field methods for the numerical data $R= \tau_{\textrm{st}}/\tau_{\textrm{mf}}$ (cyan dots) and the analytical estimate $ R= \tilde{\tau}_{\textrm{st}}/\tilde{\tau}_{\textrm{mf}}$ (black line).}
    \label{fig:relaxation_times}
\end{figure}

In the weak coupling regime ($\bar{K} \gg 1$), lattice vibrations can be treated perturbatively. Figure~\ref{fig:relaxation_times} shows the inverse relaxation time $1/\tau$ versus temperature $T$, obtained from both the stochastic master equation ($\tau_{\textrm{st}}$) and the mean-field approach ($\tau_{\textrm{mf}}$). The momentum relaxation time is extracted by initializing a Fermi-momentum wavepacket, evolving the density matrix $\rho_S$ under Eq.~\ref{eq:time evolution} with the noise $W$, and fitting the decay of Eq.~\ref{eq:expectation_value} to an exponential. In addition, perturbation theory provides analytic estimates $\tilde{\tau}_{\textrm{st}}$ and $\tilde{\tau}_{\textrm{mf}}$, detailed in Appx.~\ref{Appendix:simulation_details} and also plotted in Fig.~\ref{fig:relaxation_times}.

Figure~\ref{fig:relaxation_times} shows that the mean-field approach aligns well with the exact dynamics presented by the stochastic master equation at high temperatures (regime II). As expected in the weak coupling regime, this behavior is captured by the perturbation theory. Even though the slopes as a function of temperature agree with each other, there is a notable shift between them. This systematic deviation is caused by the zero-point fluctuations absent in the mean-field approach that induces spontaneous emission events for the system, which was also concluded in Ref.~\cite{Donghwan_phys.rev.b_106_054311_2022}. However, as seen in the inset of Fig.~\ref{fig:relaxation_times}, the stochastic and mean-field methods -- quantified either based on the simulations ($R = \tau_{\textrm{st}}/\tau_{\textrm{mf}}$) or perturbation theory ($R = \tilde{\tau}_{\textrm{st}}/\tilde{\tau}_{\textrm{mf}}$) -- approach each other relatively at high temperatures, where the temperature-independent effect of spontaneous emission is overshadowed by the increasing influence of the mean-field dynamics.

At low temperatures (regime I), both the mean-field approximation $\tau_{\textrm{mf}}$ and the stochastic master equation $\tau_{\textrm{st}}$ predict saturation of the inverse relaxation time. While $\tau_{\textrm{mf}}$ vanishes as $T \rightarrow 0$, $\tau_{\textrm{st}}$ remains finite due to zero-point lattice fluctuations inducing spontaneous emission. This temperature-independent relaxation contrasts with the mean-field contribution from lattice disorder, which scales as $\Delta V_{\textrm{def}}^2 \propto (k_BT)^{3/2}$ below the Debye temperature  $T_D$. Perturbation theory reproduces $\tau_{\textrm{mf}}$ at low $T$, but $\tilde{\tau}_{\textrm{st}}$ deviates from $\tau_{\textrm{st}}$, converging to $\tau_{\textrm{mf}}$ as $T \ll T_D$ because spontaneous emission is suppressed by fermiology. Although unphysical for metals, this effect may be relevant in dilute semiconductors.


To analyze dynamics more generally, we investigate quantum wavepacket spreading in both the strong ($\bar{K} \lesssim 1$) and weak ($\bar{K} \gg 1$) coupling regimes, particularly by benchmarking the master equation approach against the purely mean-field result. Specifically, we calculate the time-averaged spatial spread $\xi$ for the initialized wavepacket $\vert  \psi \rangle$ (for further details, see Appx.~\ref{Appendix:simulation_details}). In essence, this measure serves as a simple surrogate for transport properties of the system (see, e.g., Refs.~\cite{Donghwan_phys.rev.b_106_054311_2022, aydin_pnas_121_e2404853121_2024, aydin_pnas_122_e2426518122_2025}), while offering a practical means to test the quantum-acoustical master equation (see, e.g., Ref.~\cite{zimmermann_entropy_26_552_2024} for comparison).

\begin{figure}[h!]
    \centering
    \includegraphics[]{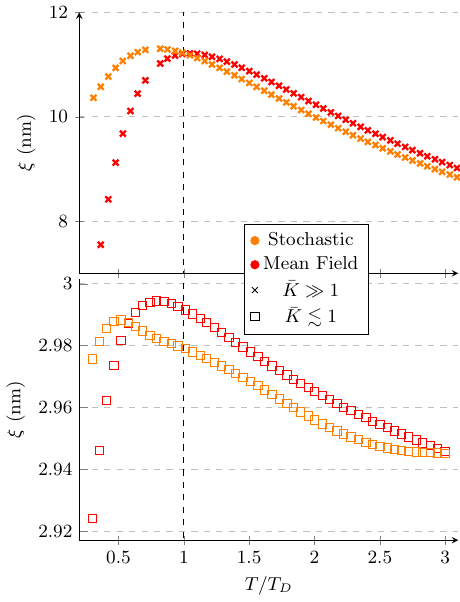}
\caption{Time-averaged spatial spread $\xi$ as a function of temperature within the weak (top) and strong (bottom) coupling domains, gauged by launching a Gaussian wavepacket with Fermi momentum. In both regimes, the spread $\xi$ is presented for the stochastic master equation (yellow) and the mean-field approach (red), with the dashed line indicating the Debye temperature of the respective material.}
\label{fig:Diffusion_coefficient}
\end{figure}

Figure~\ref{fig:Diffusion_coefficient} presents the computed spread $\xi$ as a function of temperature obtained from the full master equation (yellow) and only mean-field (red) procedures, for both the weak (top panel) and strong (bottom panel) coupling regimes. The spread lengths $\xi$ from both methods follow a similar overall trend, revealing two distinct temperature regimes separated by the Debye temperature $T_D$. 

Above this threshold ($T \gtrsim T_D$), all normal modes are thermally populated, and stimulated emission becomes the dominant process over spontaneous emission. As a result, the difference between the stochastic and mean-field methods diminishes as temperature increases. Conversely, below the Debye temperature ($T \lesssim T_D$), some lattice modes freeze out in the mean-field Hamiltonian, altering the temperature dependence of quantum spreading. In this regime, temperature-independent spontaneous emission plays a more prominent role, further amplifying the discrepancy between the stochastic and mean-field results. Interestingly, this transition occurs at a slightly lower temperature in the case of strong coupling $\bar{K} \lesssim 1$ compared to the weak coupling $\bar{K} \gg 1$. As thermal lattice vibrations are reduced at low temperatures, the zero-point fluctuations instead lead to increased quantum spreading of the stochastic results over the mean-field counterpart, which nevertheless captures the overall inclination rather well. 

We conclude that the mean-field approach is quantitatively reliable only above the Debye temperature $T/T_D \ge 1$ and loses accuracy below it. For quantitative analysis in the low-temperature regime $T/T_D \le 1$, an exact method, such as the stochastic approach presented here, is required. Nevertheless, for capturing qualitative trends, the mean-field method remains a surprisingly reliable first approximation.

We point out that there is also an alternative route for the Fr{\"o}hlich model, known as the Lee-Low-Pines transformation~\cite{lee_phys.rev_90_297_1953} if the bath is initialized in a vacuum state. This scheme of changing to the co-moving frame of the electron enables an exact treatment of electron-phonon correlations~\cite{grusdt_phys.rev.a_90_063610_2014}. Nevertheless, it also neglects phonon-phonon correlations, with the additional drawback that determining the dynamics can become cumbersome when the bath starts in a coherent or thermal state instead of a vacuum state. Beyond suffering from this impediment, our quantum-acoustical master equation accommodates any linear coupling and system Hamiltonian, without being restricted to the Fr{\"o}hlich model. 

\section{Discussion}\label{Sec:discussion}

Whereas the Fr{\"o}hlich model above served as a simple benchmark for our quantum-acoustical master equation, we outline some key differences from other approaches, such as the hierarchical equations of motion (HEOM)~\cite{tanimura_j.phys.soc.jpn_58_101_1989} and real-time path-integral methods~\cite{kundu_annu.rev.phys.chem_73_349_2022, foulkes_rev.mod.phys_73_33_2001}. These are two of the main numerical methods for simulating non-Markovian dynamics~\cite{de.vega_rev.mod.phys_89_015001_2017}. In contrast to common master equation approaches~\cite{Weiss_book, mccutcheon_phys.rev.b_84_081305_2011}, our formulation does not rely on secular, rotating-wave, or Markov approximations, while applying to arbitrary system Hamiltonians at any coupling strength. This fact makes it well-suited for capturing nonadiabatic and nonperturbative dynamics, typically inaccessible to standard master equations. Its non-Lindblad structure can be reformulated as a pair of independent Schr{\"o}dinger-like equations (see Eq.~\ref{eq:time evolution}), providing significant memory advantages.

Originally inspired by the work of Stockburger~\cite{Stockburger_vehm.phys_268_249_2001} limited to bilinear couplings and continuous spectral densities, our method extends to nonbilinear couplings, discrete baths, and the continuous state spaces relevant to the program of quantum acoustics. Earlier benchmarks of the stochastic approach for the spin–boson model~\cite{stockburger_europhys.lett_115_40010_2016, stockburger_chem.phys.296.159.2004, stockburger_phys.rev.lett_88_170407_2002} and the Fenna–Matthews–Olson (FMO) model~\cite{imai_chem.phys_446_134_2015} showed agreement with HEOM, except in one case where HEOM slightly outperformed it. However, those systems had small Hilbert spaces (2-7 sites), making HEOM viable; our interest lies in large, effectively infinite-dimensional systems, where HEOM is expected to scale poorly. In particular, the presented method is trivially parallelizable, since each noise realization is independent, unlike HEOM or path-integral methods where trajectories are interdependent. While real-time path integrals with Monte Carlo sampling share this independence, they are hampered by the dynamical sign problem~\cite{Egger_DynamicalSignProblem}. 

Whereas standard HEOM operates on density matrices whose memory cost grows rapidly with system size, wavefunction-based HEOM variants exist but are unsuitable for long-time dynamics.~\cite{Tanimura_J.Chem.Phys_153_020901} Moreover, HEOM also works best with ``well-behaved" continuous spectral densities (e.g., Lorentzian, Brownian, Ohmic)~\cite{Tanimura_J.Chem.Phys_153_020901}, and may suffer from instabilities with discrete baths~\cite{dunn_j.chem.phys_150_184109_2019}. By contrast, our quantum-acoustical master equation framework based on the wavefunction approach naturally handles discrete environments, scales well for systems near the mean-field limit, and converges quickly when bath noise is small compared to the mean-field Hamiltonian. This feature further obviates the need for regulatory methods, such as the twin-space representation~\cite{baihua_j.chem.theory.comput_21_3775_2025, zhang_j.chem.theory.comput_216352_2025}. However, we see that constrained phase-space formulations and nonadiabatic field approaches~\cite{baihua_j.chem.theory.comput_21_3775_2025, zhang_j.chem.theory.comput_216352_2025} could offer promising future directions for finite-level systems.

Because of the stochastic unraveling of the influence functional, estimations for the computational time cost are difficult. However, the stochastic Liouville-von Neumann method~\cite{Stockburger_vehm.phys_268_249_2001} on which this method is based is estimated to have $\mathcal{O}(\exp(\Gamma t))$ time scaling, where $\Gamma$ is a rate on the same scale as the decoherence rate and $t$ is the total simulation time. This can be reduced to $\mathcal{O}( t)$ by only considering finite-memory effects~\cite{stockburger_europhys.lett_115_40010_2016}. We expect the same to hold for our quantum-acoustical master equation; in the current formulation, where no approximation on the memory of the bath is made, the computational scaling is exponential. Therefore, by approximating the bath to have finite memory as in Ref.~\cite{stockburger_europhys.lett_115_40010_2016}, we expect to be able to formulate a new master equation with linear scaling.  The memory scaling is linear in the dimensions of both the system and the bath, in contrast to density matrix based methods, which have quadratic memory scaling in the system size.

We underline that the quantum-acoustical framework is not confined to our illustrative case of the Fr{\"o}hlich model above, but applies broadly to systems non-bilinearly coupled to discrete or continuous baths of harmonic oscillators $\mathcal{H}_B$. Nonetheless, it remains an open question whether a reasonable extension beyond harmonic baths is possible within the quantum-acoustical paradigm embracing the coherent-state picture. In contrast to the harmonic bath limitation, there is no restriction on the form of the system Hamiltonian $\mathcal{H}_S$. This flexibility enables applications beyond the scope of transport features in materials; for example, to study lattice effects in quantum dots, like transient localization features~\cite{KeskiRahkonen_phys.rev.lett_132_186303_2024, Ciuchi_phys.rev.B_83_081202_2011, Pustogow_nat.comm_12_1571_2021, zimmermann_entropy_26_552_2024, Ghost_of_Anderson_Arxiv, Feldmann_jacs_143_8647_2021, Blach_nano.lett_22_7811_2022, feld2026phonon}, or to investigate the emergence of thermalization~~\cite{Srednicki_phys.rev.a_50_888_1994, Deutsch_phys.rev.a_43_2046_1991, Srednicki_J.Phys.A_32_1163_1999, Rigol_Nature_452_854_2008, Rigol_Phys.Rev.Lett.108.110601_2012, Universal_quantum_birthmark_paper, Quantum_birthmark_paper} and quantum Darwinism~\cite{Zurek_Rev.Mod.Phys_75_715_2003, Zurek_Nat.Phys_5_181_2009, Zurek_book} in nanostructures. The general structure of the coupling Hamiltonian $\mathcal{H}_I$ further accommodates a broad class of interaction models, beyond the perturbative regime and the original formulation~\cite{Stockburger_vehm.phys_268_249_2001}, provided the system–bath coupling $\mathcal{H}_I$ is separable and linear in the bath coordinates $\boldsymbol{X}_{\boldsymbol{q}}$. However, more complex interactions can often be represented, or well approximated, as sums of separable terms~\cite{de.vega_rev.mod.phys_89_015001_2017}. Furthermore, our master-equation framework allows for the evaluation of arbitrary observables as formulated in Eq.~\ref{eq:expectation_value}, but also admits an extension to higher-order correlation functions, notably out-of-time-ordered correlators, thereby opening a pathway to transport calculations within the Kubo formalism (see, e.g., Ref.~\cite{KeskiRahkonen_phys.rev.lett_132_186303_2024}) as well as to fascinating investigations of quantum chaos (see, e.g., Refs.~\cite{Hashimoto_J.High.Energy.Phys_2017_1_2017, Rozenbaum_Phys.Rev.B_100_035112_2019, Jalabert_Phys.Rev.E_98_062218_2018}).
 
An important direction for future work is to develop a more complete understanding of how our master-equation framework fits within this broader setting. This will require systematic comparisons, beyond the brief discussion above, with established approaches, including related methods based on stochastic Liouville-von Neumann equations
~\cite{stockburger_phys.rev.lett_88_170407_2002, stockburger_chem.phys.296.159.2004, Schmitz_eur.phys.j_227_1929_2019, McCaul_Phys.Rev.B_95_125124_2017}, HEOM~\cite{tanimura_j.phys.soc.jpn_58_101_1989, Jin_J.Chem.Phys_128_234703_2008, Tanimura_J.Chem.Phys_153_020901, Tanimura_J.Phys.Soc.Jpn_75_082001_2006, Yan_J.Chem.Phys_140_054105_2014}, and related techniques such as Lindblad-pseudomode approaches~\cite{Pleasance_Phys.Rev.Research_2_043058_2020, Tamascelli_Phys.Rev.Lett_123_090402_2019, Garraway_Phys.Rev.A_55_2290_1997, Garraway_Phys.Rev.A.54.3592_1996, Imamoglu_Phys.Rev.A_50_3650_1994, Lentrodt_PhysRevLett.130.263602_2023, Luo_PRXQuantum_4_030316_2023, Medina_Phys.Rev.Lett_126_093601_2021}, Brownian-motion master equations~\cite{Wang_J.Chem.Phys_110_4828_1999, Shi_J.Chem.Phys_163_224103_2025, Stock_J.Chem.Phys_103_1561_1995, Shi_J.Chem.Phys_120_10647_2004, Provazza_J.Chem.Phys_151_154114_2019, Martin_J.Chem.Phys_126_164108_2007, Koch_Phys.Rev.Lett_100_230402_2008}, hierarchy-of-pure-states methods~\cite{Hartmann_J.Chem.Theory.Comput_13_5834_2017, Ritschel_J.Chem.Phys_142_034115_2015, Suess_J.Stat.Phys_159_1408_2015, Suess_Phys.Rev.Lett_113_150403_2014, Strunz_Phys.Rev.Lett_82_1801_1999, Diosi_Phys.Rev.A_58_1699_1998}, thermofield-based formulations~\cite{Diosi_Phys.Rev.A_58_1699_1998, Borrelli_New.J.Phys_22_123002_2020, Gelin_Ann.Phys_529_1700200_2017, Gelin_J.Chem.Phys_155_134102_2021, Yu_Phys.Rev.A_69_062107_2004, de.Vega_Phys.Rev.A_92_052116_2015}, and non-Markovian perturbative treatments~\cite{de.vega_rev.mod.phys_89_015001_2017, Breuer_Rev.Mod.Phys_88_021002_2016, StockburgerColloquium, Gonzalez_Quantum_8_1454_2024, Mulvihill_J.Phys.Chem.B_125_9834_2021}. For instance, while our formulation assumes an initially factorized state, the ensuing dynamics can generate strong system-bath entanglement, unlike mean-field approaches, but this feature is not fully resolved in the Fr{\"o}hlich benchmark, and hence warrants further study. Similarly, it has yet to be demonstrated whether the present approach can (competitively) capture topological effects, such as geometric phases, particularly in light of recent advances in the multi-state path-integral approach~\cite{Liu_J.Chem.Phys_145_024103_2016, Zhang_J.Chem.Phys_034109_2017, Liu_J.Chem.Phys_148_102319_2018, Zhai_J.Phys.Chem.Lett_17_4274_2026}.

Finally, we reiterate that the primary ambition behind this work is to further develop the emerging program of quantum acoustics, rather than to introduce a superior numerical scheme. At the same time, the present framework of the quantum-acoustical master equation provides a numerically exact, fully quantum, non-Markovian description for the time evolution of a generic system coupled to a harmonic bath, akin to stochastic Liouville-von Neumann equation approaches in spirit~\cite{stockburger_phys.rev.lett_88_170407_2002, stockburger_chem.phys.296.159.2004, Schmitz_eur.phys.j_227_1929_2019, McCaul_Phys.Rev.B_95_125124_2017} but building on our refinement of Stockburger’s methodology~\cite{Stockburger_vehm.phys_268_249_2001, StockburgerColloquium}. While we have discussed a preliminary numerical implementation, its overall performance depends on the selected solution strategy, and systematic optimization lies beyond the scope of this work. Instead, we have established internal consistency and validated the concept in controlled settings, with the observed features, such as the reported rapid convergence, indicating promising directions for future development.  

\section{Outlook and conclusion}

In general, the quantum-acoustical stochastic master equation provides an alternative view of lattice baths, while also enabling studies of strong, nonperturbative, nonadiabatic electron–lattice interactions. Beyond strange metals, it is suitable for materials with low-lying or tunable Fermi levels including perovskites~\cite{han_energy.mater_5_0116_2024, lafalce_acs.nano_18_18299_2024}, dilute semiconductors~\cite{lin_phys.rev.x_3_021002_2013, skipetrov_j.alloys.compd_893_162330_2022}, and twisted bilayer graphene~\cite{nimbalkar_nanomicro.letter_12_1_2020, cea_phys.rev.b_100_205113_2019, zhang_nat.commun_15_3737_2024}. The framework can help to understand Planckian transport~\cite{aydin_pnas_121_e2404853121_2024, Varma_rev.mod.phys_92_031001_2020, Bruin_science_339_804_2013,Polshyn_nat.phys_15_1011_2019}, frustrated Anderson localization~\cite{KeskiRahkonen_phys.rev.lett_132_186303_2024, Ciuchi_phys.rev.B_83_081202_2011, Pustogow_nat.comm_12_1571_2021, zimmermann_entropy_26_552_2024, Ghost_of_Anderson_Arxiv}, Mott–Ioffe–Regel saturation~\cite{hussey_philos.mag._84_2847_2004, gunnarsson_rev.mod.phys_75_1085_2003}, and displaced Drude peaks~\cite{KeskiRahkonen_phys.rev.lett_132_186303_2024, Biswas_phys.rev.lett_125_076403_2020, Pustogow_nat.comm_12_1571_2021, Jaramillo_nat.phys_10_304_2014, Michon_nat.comm_14_3033_2023}, and opens avenues for exploring dynamical polaron formation~\cite{aydin_pnas_122_e2426518122_2025, Vasiliu_phys.rev.lett_83_4393_1999,Guzelturk_nat.mater_20_618_2021} and localization-enhanced photon yield in quantum dots~\cite{Feldmann_jacs_143_8647_2021, Blach_nano.lett_22_7811_2022, feld2026phonon}. As illustrated by some of the aforementioned examples, we note that the time-dependent fields and potentials central to quantum acoustics virtually call for wavepacket-based approaches.

The quantum-acoustical framework has mainly addressed coherent baths with discrete spectra~\cite{Donghwan_phys.rev.b_106_054311_2022, KeskiRahkonen_phys.rev.lett_132_186303_2024, aydin_pnas_121_e2404853121_2024, aydin_pnas_122_e2426518122_2025, zimmermann_entropy_26_552_2024}, but outside of this framework coherent lattice baths have garnered relatively little attention. Our stochastic path-integral approach naturally incorporates such baths, overcoming limitations of many existing techniques~\cite{de.vega_rev.mod.phys_89_015001_2017} that often rely on Markovian approximations, fail to capture nonperturbative real-space dynamics, or struggle with large discrete baths required for realistic materials, like in the case of the Fr{\"o}hlich model. A future line of inquiry includes properly benchmarking its computational efficiency against other comparable methods (see Sec.~\ref{Sec:discussion}). In fact, an important direction for future work is a comprehensive assessment of the capabilities and limitations of the quantum-acoustical master equation.

On the theoretical frontier, the stochastic master equation can be advanced by introducing norm-preserving noise~\cite{Stockburger_vehm.phys_268_249_2001}, variance reduction~\cite{Schmitz_eur.phys.j_227_1929_2019}, or other initial states with an additional path integral over imaginary time~\cite{Weiss_book}. For instance, one promising avenue is to extend the present implementation by generalizing the wavepacket to be thermal, as described in Ref.~\cite{heller_nano.lett_5_1285_2005}, while simultaneously improving the associated master-equation propagator to enhance computational speed, accuracy, and memory efficiency. Further extensions include incorporating second-order acoustic coupling and optical modes via an optical deformation potential~\cite{Mahan_book}. More broadly, this framework may bridge toward path-integral Monte Carlo methods~\cite{Herrero_j.condens.matter.phys_26_233201_2014} and connect with density functional theory, where the deformation potential approach is well established~\cite{Li_phys.rev.B_104_195201_2021, Resta_phys.rev.b_44_11035_1991, Jin_npj_9_190_2023}.

In summary, we introduced a path-integral framework for lattice vibrations within the coherent state formalism that solidifies quantum acoustics as a parallel program to quantum optics. In particular, we have set forth a quantum-acoustical, stochastic master equation describing the exact dynamics for any system linearly coupled to a harmonic lattice, whose potential we illustrated for the traditional Fr{\"o}hlich model. More generally, we have established a quantum-acoustical avenue for shedding light on strong electron-lattice dynamics.


The authors thank Alhun Aydin and Nikolai Leopold for fruitful discussions and suggestions. This work was supported by the U.S. Department of Energy under Grant No. DE-SC0025489. A.M.G. thanks the Studienstiftung des Deutschen Volkes for financial support. J.K.-R. thanks the Oskar Huttunen Foundation for financial support. 

\appendix

\section{Quantum-acoustical master equation}\label{Appendix:master_equation}

Here we derive the quantum-acoustical master equation for a generic system $\mathcal{H}_S$ that is linearly coupled to a bath of harmonic lattice vibrations $\mathcal{H}_B$. Our master plan reflects the stochastic method by Stockburger\textit{et al.}~\cite{Stockburger_vehm.phys_268_249_2001} that we utilize to institute a mathematically rigorous foundation for the emerging field of quantum acoustics. However, we here provide an extension to this framework, as the case of non-bilinear coupling with discrete or coherent baths has not been explored before.

\subsection{System setup}

In this work, we consider the following composite system
\begin{equation}
    \mathcal{H}=\mathcal{H}_S+\mathcal{H}_B+\mathcal{H}_I.
\end{equation}
In other words, a general (unspecified) system characterized by the Hamiltonian $\mathcal{H}_s$ interacts with a lattice bath $\mathcal{H}_B$ assumed to be instituted by a set of independent harmonic oscillators with frequencies $\omega_{\mathbf{q}}$, or formally
\begin{equation}
    \mathcal{H}_B = \frac{1}{2}\sum_{\boldsymbol{q}} \omega_{\boldsymbol{q}}\Big (\boldsymbol{X}_{\boldsymbol{q}}\cdot\boldsymbol{X}_{\boldsymbol{q}}^T \Big),
\end{equation}
where the variable $\boldsymbol{X}_{\boldsymbol{q}} =\left[\begin{matrix} x_{\boldsymbol{q}} & p_{\boldsymbol{q}}  \end{matrix}\right]$ collectively denotes the position $x_{\boldsymbol{q}}$ and momentum $p_{\boldsymbol{q}}$ of normal mode $\boldsymbol{q}$ of lattice vibrations. The coupling between the system and bath is assumed to be linear in the most general way, translating to
\begin{equation}\label{Eq_app:interaction_Hamiltonian}
    \mathcal{H}_I = \frac{1}{2}\sum_{\boldsymbol{q}} \Big( \boldsymbol{g}_{\boldsymbol{q}}(\boldsymbol{r}) \cdot \boldsymbol{X}_{\boldsymbol{q}}^T \Big),
\end{equation}
where $\bm g_{\bm q}(\bm r) = \left[\begin{matrix} g^0_{\bm q}(\bm r) & g^1_{\bm q}(\bm r) \end{matrix}\right]$ are functions depending only on the system coordinates $\boldsymbol{r}(t)$. These coupling functions $g^0_{\bm q}(\bm r)$ and $g^1_{\bm q}(\bm r)$ are undetermined for us and are specified by the given system, thus highlighting the general nature of our master equation presented below. Notably we have no requirement for the dispersion $\omega(\boldsymbol{q})$ of the lattice modes. 

\subsection{Initial state of the system}

We start by assuming the initial density matrix $\rho$ of the composite system is separable, i.e., 
\begin{equation}\label{eq_app:initial_condition}
\hat \rho(t=0)=\hat\rho_S\otimes \hat\rho_B,
\end{equation}
where $\rho_S$ and $\rho_B$ are the individual density matrices for the studied system and bath at time $t = 0$, respectively. This assumption is a common starting point~\cite{Weiss_book}, disentangling the system and bath initially. However, the composite system is not restricted to stay in the direct form throughout its time evolution within the presented quantum-acoustical formalism, contrast to mean-field approaches, such as those presented in Refs.~\cite{Donghwan_phys.rev.b_106_054311_2022, aydin_pnas_121_e2404853121_2024,aydin_pnas_122_e2426518122_2025}.

Within the coherent framework~\cite{Donghwan_phys.rev.b_106_054311_2022}, each normal mode of lattice vibration with a wave vector $\mathbf{q}$ is associated with a coherent state $\vert \alpha_{\mathbf{q}} \rangle$. Employing the independence of normal modes, the total set of lattice vibrations can be described as the product state of the coherent states of the normal modes ~\cite{heller_J.Phys.Chem.A_123_4379_2019}, i.e., as a multimode coherent state $\vert \bm{\alpha} \rangle = \prod_{\mathbf{q}}\vert\alpha_{\mathbf{q}} \rangle$. By introducing the notation
\begin{equation}
    \bar{\boldsymbol{A}}
    = \begin{bmatrix}
        1 & 0\\
        0 & 0
    \end{bmatrix}
    \quad \textrm{and} \quad
    \bar{\boldsymbol{Z}}
    = \begin{bmatrix}
        0 & 0\\
        i & 0
    \end{bmatrix},
\end{equation}
we can write down the corresponding initial density matrix of the bath as
\begin{equation}
\label{eq_app:coherent bath}
\begin{split}
    \rho_B( \boldsymbol{X},\boldsymbol{X}';\boldsymbol{X}^0) \\= \mathcal{N} \exp \Bigg\{&-\frac{1}{2} \sum_{\boldsymbol{q}} \Big[ (\boldsymbol{X}_{\boldsymbol{q}}  - \boldsymbol{X_{\boldsymbol{q}} }^0)  \cdot \bar{\boldsymbol{A}} \cdot ( \boldsymbol{X}_{\boldsymbol{q}}  - \boldsymbol{X}_{\boldsymbol{q}} ^0)^T\\ &+ (\boldsymbol{X}_{\boldsymbol{q}}' - \boldsymbol{X}_{\boldsymbol{q}}^0)  \cdot \bar{\boldsymbol{A}} \cdot ( \boldsymbol{X}_{\boldsymbol{q}}' - \boldsymbol{X}_{\boldsymbol{q}}^0 )^T\\ &-2\boldsymbol{X}_{\boldsymbol{q}}^0\cdot \bar{\boldsymbol{Z}} \cdot ( \boldsymbol{X}_{\boldsymbol{q}} - \boldsymbol{X}_{\boldsymbol{q}}' )^T \Big] \Bigg \},
\end{split}
\end{equation}
where $\mathcal{N} = \pi^{-N/2}$ is a normalization constant and $\bm{X^0}$ are the expected values of the coordinates at $t=0$.

Instead of taking the single multimode coherent state above, a thermal ensemble can also determined for the initial lattice bath similar to quantum  optics~\cite{gauber_phys_rev_131_2766_1963} as
\begin{equation}\label{eq_app:thermal prob distr}
\hat \rho_T=\int P( \boldsymbol{X}^0)\hat\rho_B( \boldsymbol{X}^0)d \boldsymbol{X}^0
\end{equation}
with respect to the following probability distribution 
\begin{equation*}
    P(\boldsymbol{X})=\prod_{\bm q} \frac{1}{\pi \bar{n}_{\bm q}}\exp\left[ -\frac{\boldsymbol{X}_{\boldsymbol{q}}\cdot\boldsymbol{X}_{\boldsymbol{q}}}{2\bar{n}_{\bm q}} \right],
\end{equation*}
with
\begin{equation*}
    \bar{n}_{\bm q}=\frac{1}{\exp( \hbar\omega_{\bm q}/ k_B T)-1},
\end{equation*}
where $\bar{n}_{\bm q}$ is the average number of quanta in a mode given by the Bose-Einstein distribution. However, at high temperatures, this distribution is sharply peaked around the Bose-Einstein value $\bar{n}_{\bm q}$. In this case,  thermal coherent state $\vert \alpha_\mathbf{q} \rangle$ can be approximated as 
\begin{equation}
    \alpha_\mathbf{q}=\sqrt{ \bar{n}_{\bm q}}\exp(i\varphi_\mathbf{q}),
\end{equation}
where $\sqrt{\bar{n}_{\bm q}}=\vert\alpha_\mathbf{q}\vert$ is the thermal amplitude and $\varphi_{\mathbf{q}}=\arg(\alpha_{\mathbf{q}})$ is the random phase of the coherent state determining the initial conditions. This approach has been successfully applied to lattice vibrations (see, e.g. Refs.~\cite{Donghwan_phys.rev.b_106_054311_2022, aydin_pnas_121_e2404853121_2024, KeskiRahkonen_phys.rev.lett_132_186303_2024}). Later, we only consider the multimode coherent state of Eq.~\ref{eq_app:coherent bath}, instead of the thermal one of Eq.~\ref{eq_app:thermal prob distr}. Nevertheless, the following derivation of the quantum-acoustical master equation holds for the thermal initial condition of Eq.~\ref{eq_app:thermal prob distr}. 

\subsection{Defining the influence functional}

Let $\bm r$ and $\bm x$ denote the system and bath coordinates, respectively. Then the reduced density matrix of the studied system can be expressed~\cite{Weiss_book} in the form of
\begin{equation}
\rho_S(\bm{r}_f,\bm{r}'_f;t)=\int d\bm{r}_id\bm{r}'_i J(\bm{r}_f,\bm{r}_f',t;\bm{r}_i,\bm{r}_i',0)\rho_S(\bm{r}_i,\bm{r}'_i;0),
\end{equation}
where the subscripts $i$ and $f$ stand for initial and final, respectively and $J$ is the propagator for the density matrix. The propagator $J$ is on the other hand defined by a path-integral~\cite{Feynman_ann.phys_24_118_1963} as
\begin{equation}\label{Eq_app:propagator}
\begin{split}
&J(\bm{r}_f,\bm{r}_f',t;\bm{r}_i,\bm{r}_i',0)\\&=\int\mathcal{D}\boldsymbol{r}\mathcal{D}\boldsymbol{r}'\exp(i S_S[\bm r]-iS_S[\bm r'])\mathcal{F}[\bm r,\bm r'],
\end{split}
\end{equation}
with $S_S$ and $\mathcal{F}$ denoting the system, action and the influence functional, respectively. The influence functional incorporates the contribution of the bath to the system, defined as 
\begin{align}
\begin{split}
    \mathcal{F}[\bm q,\bm q']&=\int d\bm{x}_fd\bm{x}_id\bm{x}'_i \rho_B(\bm x_i,\bm x'_i)\\& \times \int\mathcal{D}\bm x\mathcal{D} \bm x'\exp\Big[i(S_B[\bm x]+S_I[\bm x,\bm q]\\&-S_B[\bm x']-S_I [\bm x',\bm q']) \Big],
\end{split}
\end{align}
where $S_B$ and $S_I$ are the actions for the bath and system, respectively.

For our system,  the influence functional can be computed as
\begin{equation}\label{Eq_app:functional_trace_form}
\mathcal{F}[\bm r,\bm r']=\text{Tr}_B\left(\hat\rho_B \hat U^{\dagger}_B[\bm r',t]\hat U_B[\bm r,t]\right),
\end{equation}
employing the density operator of the bath $\hat\rho_B$  and its time evolution operator $\hat U^{\dagger}_B[\bm r',t]$. In the interaction picture, we can write
\begin{align}
\begin{split}
\mathcal{H}_I(t)&=\exp \left({i\mathcal H_Bt} \right)
\mathcal{H}_I(t) \exp \left(-i\mathcal H_Bt\right)\\&=\sum_{\boldsymbol{q}} \hat{x}_{\boldsymbol{q}}\left[g^0_{\boldsymbol{q}}(t)\cos(\omega_{\boldsymbol{q}}t)-g^1_{\boldsymbol{q}}(t)\sin(\omega_{\boldsymbol{q}}t) \right]\\&+\hat{p}_{\boldsymbol{q}} \left[g^0_{\boldsymbol{q}}(t)\sin(\omega_{\boldsymbol{q}}t)+g^1_{\boldsymbol{q}}\cos(\omega_{\boldsymbol{q}}t)\right]
\end{split}
\end{align}
The propagator of the bath in the interaction picture is given by 
\begin{equation}
\label{eq_app:Schrodinger equation}
    i\partial_t U_B(t)=\mathcal{H}_I(t)U_B(t),
\end{equation}
which can be solved by taking the following ansatz:
\begin{equation}
    U_B(t)=e^{-iA(t)}e^{-i\sum_{\boldsymbol{q}} B_{\boldsymbol{q}}(t) \hat{x}_{\boldsymbol{q}}}e^{-i\sum_{\boldsymbol{q}}C_{\boldsymbol{q}}(t)\hat{p}_{\boldsymbol{q}}}.
\end{equation}
with the following coefficients 
\begin{equation}
\begin{split}
A_{\boldsymbol{q}}=&-\int_0^t\int_0^{t'}(g^0_{\boldsymbol{q}}(s)\cos(\omega_{\boldsymbol{q}}s)-g^1_{\boldsymbol{q}}(s)\sin(\omega_{\boldsymbol{q}}s))\\ &\times (g^0_{\boldsymbol{q}}(t')\sin(\omega_{\boldsymbol{q}}t')+g^1_{\boldsymbol{q}}(t')\cos(\omega_{\boldsymbol{q}}t'))dsdt',
\end{split}
\end{equation}
\begin{equation}
B_{\boldsymbol{q}}(t)=\int_0^t g^0_{\boldsymbol{q}}(t')\cos(\omega_{\boldsymbol{q}}t')-g^1_{\boldsymbol{q}}(t')\sin(\omega_{\boldsymbol{q}}t')dt',
\end{equation}
\begin{equation}
C_{\boldsymbol{q}}(t)=\int_0^t g^0_{\boldsymbol{q}}(t')\sin(\omega_{\boldsymbol{q}}t')+g^1_{\boldsymbol{q}}(t')\cos(\omega_{\boldsymbol{q}}t')dt',
\end{equation}
Therefore we can write the influence functional in Eq.~\ref{Eq_app:functional_trace_form} as
\begin{align}
\begin{split}
\label{eq_app:influence functional calculation}
    &\mathcal{F}[\bm r(t'),\bm r'(t')]
    \\ &=\text{Tr}_B\big(\rho_B U^{\dagger}_B[\bm r']U_B[\bm r]\big)\\&
=e^{i\sum_{\boldsymbol{q}} (A_{\boldsymbol{q}}'-A_{\boldsymbol{q}})}\int d\boldsymbol{x} \langle \boldsymbol{x} |\rho_B e^{i\sum_{\boldsymbol{q}}C_{\boldsymbol{q}}'\hat{p}_{\boldsymbol{q}}}\\ &\times e^{i\sum_{\boldsymbol{q}}(B_{\boldsymbol{q}}'-B_{\boldsymbol{q}})\hat{x}_{\boldsymbol{q}}}e^{-i\sum_{\boldsymbol{q}}C_{\boldsymbol{q}}\hat{p}_{\boldsymbol{q}}}|\boldsymbol{x}\rangle\\&
= e^{i\sum_{\boldsymbol{q}} (A_{\boldsymbol{q}}'-A_{\boldsymbol{q}})}\int d\boldsymbol{x} \langle \boldsymbol{x} |\rho_B |\boldsymbol{x}-\boldsymbol{C} +\boldsymbol{C}'\rangle\\& \times e^{i\sum_{\boldsymbol{q}}(B_{\boldsymbol{q}}'-B_{\boldsymbol{q}})(x_{\boldsymbol{q}}-C_{\boldsymbol{q}})},
\end{split}
\end{align}
where we have denoted $|\boldsymbol{x}\rangle=\prod_{\boldsymbol{q}} |x_{\boldsymbol{q}}\rangle$ and $|\boldsymbol{x}-\boldsymbol{C}+\boldsymbol{C'}\rangle=\prod_{\boldsymbol{q}} |x_{\boldsymbol{q}}-C_{\boldsymbol{q}}+C_{\boldsymbol{q}}'\rangle$.
Moreover, when our initial bath state is Gaussian as assumed in Eq.~\ref{eq_app:coherent bath}, the influence functional is given by a Gaussian integral, which can be evaluated to be
\begin{equation}
    \mathcal{F}[\bm r(t'),\bm r'(t')]=\exp{[-\Phi[\bm r(t'),\bm r'(t')]},
\end{equation}
where we have defined the phase factor as
\begin{align*}
    &\Phi[\xi(t'),\xi'(t')]\\
    &=-\Big[i (A'-A)-i\boldsymbol{p}_0\cdot (\boldsymbol{C}-\boldsymbol{C}')
-\frac{1}{4}(\boldsymbol{C}-\boldsymbol{C}')^2\\&-\frac{1}{4}(\boldsymbol{B}'-\boldsymbol{B})^2+i(\boldsymbol{B}'-\boldsymbol{B})\cdot\boldsymbol{x}_0\\ &-\frac{1}{2}i(\boldsymbol{C}+\boldsymbol{C}')\cdot(\boldsymbol{B}'-\boldsymbol{B})\Big].
\end{align*}

On the other hand, we can construct the mean-field action as
\begin{align*}
&S_{\text{mf}}[\bm r(t'),\bm r'(t')]=-\sum_{\bm q}\int_0^t\\&{x}_{0, \bm q} \left(g^0_{\boldsymbol{q}}(t')\cos(\omega_{\boldsymbol{q}}t')-g^1_{\boldsymbol{q}}(t')\sin(\omega_{\boldsymbol{q}}t')\right) +\\&{p}_{0,\bm q}\left(g^0_{\boldsymbol{q}}(t')\sin(\omega_{\boldsymbol{q}}t')+g^1_{\boldsymbol{q}}(t')\cos(\omega_{\boldsymbol{q}}t')\right)dt',
\end{align*}
that enables the subsequent decomposition of the influence functional 
\begin{equation}
\begin{split}
\mathcal{F}[\bm r(t),\bm r'(t)] &=\exp  \Big\{-iS_{\text{mf}}[\bm r'(t)]+iS_{\text{mf}}[\bm r(t)]\\  &-\Psi[\bm r(t),\bm r'(t)] \Big\},
\end{split}
\end{equation}
where $\Psi$ is the influence phase
\begin{equation}
\begin{split}
    \Psi[\xi(t'),\xi'(t')]&=i (A-A')+\frac{1}{4}(\boldsymbol{C}-\boldsymbol{C}')^2\\ &+\frac{1}{4}(\boldsymbol{B}'-\boldsymbol{B})^2+\frac{1}{2}i(\boldsymbol{C}+\boldsymbol{C}')\cdot(\boldsymbol{B}'-\boldsymbol{B}).
\end{split}
\end{equation}
Notably, since the phase $\Psi$ is quadratic in the coupling $g$, the mean-field component $S_{\text{mf}}$ that is instead linear dominates the system behavior  as $g\rightarrow0$. In other words, the common mean-field approximation is well justified in this limit.

As a final step, we move into the center of mass $\bm u$ and fluctuation coordinates $\bm v$ specified as
\begin{equation}\label{Eq_app:new_coordinates}
\begin{split}
    \bm u_{\bm q}(t)&=\frac{1}{2}\left[\bm g_{\bm q}(\bm r,t)+\bm g_{\bm q}(\bm r',t)\right],\\ \\
    \bm v_{\bm q}(t)&=\bm g_{\bm q}(\bm r,t)-\bm g_{\bm q}(\bm r',t),
\end{split}
\end{equation}
as well as call upon the following functions
\begin{equation}
\begin{split}
    \bar{\boldsymbol{L}}_{\bm q}(t'-s)=
    \begin{bmatrix}
    \cos[\omega_{\bm q}(t'-s)] &-\sin[\omega_{\bm q}(t'-s)] \\\sin[\omega_{\bm q}(t'-s)] &\cos[\omega_{\bm q}(t'-s)] 
    \end{bmatrix},\\\\
    \bar{\boldsymbol{M}}_{\bm q}(t'-s)=
    \begin{bmatrix}\sin[\omega_{\bm q}(t'-s)] &0 \\0 &-\sin[\omega_{\bm q}(t'-s)]
    \end{bmatrix}.
\end{split}    
\end{equation}
Within this convention, we express the mean-field in the compact manner as
\begin{equation*}
    S_{\text{mf}}[\bm r(t')]=-\sum_{\bm q}\int_0^t \boldsymbol{X}_{\bm q}^0\cdot\bar{\boldsymbol{L}}_{\bm q}(t') \cdot\boldsymbol{g}_{\bm q}(\boldsymbol{r'}) dt',
\end{equation*}
along with the influence phase
\begin{align*}
\Psi[\bm r(t),\bm r'(t)]=&
-\frac{1}{2}\int_0^t\int_0^t\sum_{\bm q}\\& \frac{1}{2}\bm v_{\bm q}(s)\cdot \bar{\boldsymbol{L}}_{\bm q}(t'-s)\cdot \bm v_{\bm q}(t')+ \\&
2i\bm v_{\bm q}(s)\cdot \bar{\boldsymbol{M}}_{\bm q}(t'-s)\Theta(s-t')\cdot \bm u_{\bm q}(t')dt'ds.
\end{align*}
As we have computed the mean-field $S_{\text{mf}}$ and influence phase $\Psi$, the propagator $J$ is technically solved, thus determining the reduced density matrix $\rho_S$ of the studied system from the given initial condition $\hat{\rho}(t=0)$ in Eq.~\ref{eq_app:initial_condition}. 

\subsection{Master equation}

Subsequently, we can establish a stochastic master equation for the density matrix $\hat{\rho}_S$ of a pseudo-potential $\mathcal{H}_{B+I}^{\pm}$, which recovers the original density matrix $\rho_S$  when averaged over the prescribed noise $W$, i.e., $\rho_S=\left<\tilde\rho_S\right>_W$, as derived below.

This goal is achieved by puting in place artificial noise terms $\bm \eta(t)$ and $\bm \nu(t)$ that are normally distributed with zero mean and covariances given by
\begin{equation} \label{eq_app:noise_covariance_original}
\begin{split}
\Big \langle \bm \eta^T_{\bm q}(s) \cdot \bm \eta_{\boldsymbol{r}}(t')\Big \rangle_W &=\frac{1}{2} \bar{\boldsymbol{L}}_{\bm q}(t'-s)\delta_{\bm q,\boldsymbol{r}},
\\
\\
\Big \langle \bm\eta^T_{\boldsymbol{q}}(s)\bm \cdot \bm \nu_{\bm r}(t')\Big \rangle_W &=i\bar{\boldsymbol{M}}_{\bm q}(t'-s)\Theta(s-t')\delta_{\bm q,\boldsymbol{r}},
\\
\\
\Big \langle \bm\nu^T_{\bm q}(s) \cdot \bm\nu_{\boldsymbol{r}}(t')\Big \rangle_W &=0.
\end{split}
\end{equation}
Then by utilizing the Hubbard-Stratonovich transformation~\cite{Hubbard_phys.rev.lett_3_77_1959, stratonovich_book}, we can transform the influence phase into
\begin{equation}
\begin{split}
    \Psi[\bm r(t),\bm r'(t)] &= - \ln \Big[ \Big<\exp\big(i\int_0^t\sum_{\bm q} \bm{\eta}_{\bm q}(s)\cdot \bm{v}_{\bm q}(s)\\ &+\bm{\nu}_{\bm q}(s)\cdot\bm{u}_{\bm q}(s)\,ds \big)\Big>_{W} \Big],
\end{split}    
\end{equation}
which further converts into
\begin{equation}
\begin{split}
    &\exp(\Psi[\bm r(t),\bm r'(t)])\\&=\Bigg<\exp\Big(-i\int_0^t\sum_{\bm q} \bm{\eta}_{\bm q}(s)\cdot (\bm g_{\bm q}(\bm r,s)-\bm g_{\bm q}(\bm r',s)\big)\\ &+\bm{\nu}_{\bm q}(s)\cdot\frac{1}{2}\left(\bm g_{\bm q}(\bm r,s)+\bm g_{\bm q}(\bm r',s)\right) ds \Big)\Bigg>_{W} 
\end{split}    
\end{equation}
when substituting the definitions for the center-of-mass $\bm u$ and fluctuation coordinates $\bm v$ defined in Eq.~\ref{Eq_app:new_coordinates}. We thus see that we can define the corresponding noise actions as
\begin{equation}
    S_{\bm \eta}[\bm r]=-\int_0^t\sum_{\bm q}\bm \eta_{\bm q} \cdot \bm g_{\bm q}(\bm r,s)ds
\end{equation}
and
\begin{equation}
    S_{\bm \nu}=-\frac{1}{2}\int_0^t\sum_{\bm q}\bm \nu_{\bm q} \cdot\bm g_{\bm q}(\bm r,s)ds
\end{equation}
resulting in the influence phase expansion of
\begin{equation}
\begin{split}
        &\exp(\Psi[\bm r(t),\bm r'(t)])\\ &=\left<\exp\left(i S_{\bm \eta}[\bm r]-iS_{\bm \eta}[\bm r']+iS_{\bm \nu}[\bm r]+iS_{\bm \nu}[\bm r']\right)\right>_{W}.
\end{split}        
\end{equation}

In other words, the propagator $J$ for the system density matrix $\rho_S$ stated in Eq.~\ref{Eq_app:propagator} takes the form of
\begin{equation}
\begin{split}
&J(\bm{r}_f,\bm{r}_f',t;\bm{r}_i,\bm{r}_i',0)\\&=\int\mathcal{D}\boldsymbol{r}\mathcal{D}\boldsymbol{r}'\exp(i S_S[\bm r]-iS_S[\bm r'])\mathcal{F}[\bm r,\bm r']\\&=\int\mathcal{D}\boldsymbol{r}\mathcal{D}\boldsymbol{r}'\exp(i S_S[\bm r]-iS_S[\bm r']+iS_{\text{mf}}[\bm r]\\&-iS_{\text{mf}}[\bm r']-\Psi[\bm r,\bm r'])\\
&=
\int\mathcal{D}\boldsymbol{r}\mathcal{D}\boldsymbol{r}' \left<\exp\left(i S_S[\bm r]-iS_S[\bm r']+iS_{\text{mf}}[\bm r]-iS_{\text{mf}}[\bm r']\right.\right.\\&
\left.\left.-i S_{\bm \eta}[\bm r]+iS_{\bm \eta}[\bm r']-iS_{\bm \nu}[\bm r]-iS_{\bm \nu}[\bm r']\right)\right>_{W}.
\end{split}
\end{equation}
Based on the construction of the action above, we see that the reduced density matrix $\tilde\rho_S$ for a single noise realization follows a stochastic Liouville-von Neumann equation
\begin{equation}
\label{eq_app:equation_of_motion_1}
\begin{split}
    i\frac{\partial}{\partial t}\tilde{\rho}_S &= \Big[\mathcal{H}_{S}+\mathcal{H}_{\text{mf}},\tilde\rho_S \Big]\\ &- \sum_{\boldsymbol{q}} \left( [\bm\eta_{\boldsymbol{q}}(t)\cdot \bm g_{\boldsymbol{q}},\tilde \rho_S]-\frac{1}{2}\{\bm\nu_{\boldsymbol{q}}(t)\cdot \bm g_{\boldsymbol{q}},\tilde \rho_S\} \right).
\end{split}
\end{equation}
Here, we want to highlight that the derived master equation is valid beyond the weak system-bath coupling and the Markovian approximation. Nevertheless, we see that the mean-field approximation becomes more accurate in the weak-coupling limit, since the mean-field action is linear in the coupling whereas the influence phase is quadratic. Moreover, unlike Lindbladian approaches~\cite{Weiss_book}, our quantum-acoustical master equation can be further decomposed into two independent Schr{\"o}dinger equations with non-Hermitian Hamiltonians. By adopting the initial state of $\rho_S(t=0)=|\psi_{+}\rangle\langle\psi_{-}|$, we can then track the evolution of these states as
\begin{equation}\label{eq_app:equation_of_motion_2}
\begin{split}
    i \frac{\partial}{\partial t}|\psi_{\pm}\rangle =\Bigg(\mathcal{H}_{S}+\mathcal{H}_{B+I}^{\pm}\Bigg)|\psi_{\pm}\rangle,
\end{split}   
\end{equation}
where we define a stochastic pseudopotential that encapsulates the lattice interaction within the system as

\begin{equation}
\begin{split}
    \mathcal{H}_{B+I}^{\pm} = \left[\mathcal{H}_{\textrm{mf}} - \sum_{\bm q} \Big(\bm \eta_{\bm q}(t)\cdot \bm g_{\bm q} \mp \frac{1}{2}\bm \nu_{\bm q}(t)\bm \cdot \bm g_{\bm q} \Big)\right]^{-/\dagger}. 
\end{split}   
\end{equation}

In essence, implementing the quantum-acoustical master equation framework involves first specifying the system Hamiltonian $\mathcal{H}_S$ and defining the interaction via the coupling functions $\bm g_{\bm q}$. Next, one initializes the system state $\hat{\rho}(t= 0)$ along with the coherent state parameters $\bm X_{\bm q}$. The evolution of the system state $\tilde{\rho}_S$ for a given noise realization is governed by Eq.~\ref{eq_app:equation_of_motion_1}, with Gaussian noise $W$ sampled in accordance with Eq.~\ref{eq_app:noise_covariance_original}, and the physical density matrix emerges as averaged over the disorder $W$. Subsequently, the expectation value of any system observable $\hat O(t)$ can be evaluated as
\begin{equation}
    \langle\hat O \rangle (t) = \textrm{Tr}\big[ \rho_S \hat{O} \big]
    \quad \textrm{with} \quad
    \rho_S=\left<\tilde\rho_S\right>_W.
\end{equation}
Alternatively, one can initialize the states $|\psi_{\pm}\rangle$ and determine their time evolution according to Eq.~\ref{eq_app:equation_of_motion_2}, and the expectation value of the operator $\hat{O}$ is given by
\begin{equation}
    \langle\hat O \rangle (t) =\left<\langle\psi_-(t)|\hat O|\psi_+(t) \rangle\right>_W.
\end{equation}
Furthermore, in the following section, we illustrate this procedure in action with an explicit Fr{\"o}hlich model. In addition, we delve deeper into the numerical techniques to efficiently utilize the developed quantum-acoustical master equation.

\section{Simulation details}\label{Appendix:simulation_details}

As an example, we consider the standard Fr{\"o}hlich model~\cite{Frochlich_adv.phys_3_325_1954, Frochlich_proc.r.soc.a_160_230_1937} where the Hamiltonian is composed of the sum of the electronic band energy (first term), the elastic energy of lattice vibrations (second term) and their interaction energy (third term) in the Schr{\"o}dinger picture~\cite{Mahan_book} which is given as follows
\begin{equation}\label{Eq_app:Frohlich_Hamiltonian}
\begin{split}
    \mathcal{H} &= \sum_{\boldsymbol{p}}\varepsilon(\boldsymbol{k}) c_{\mathbf{p}} c_\mathbf{p}^{\dagger} + \sum_{\boldsymbol{q}} \hbar \omega_{\boldsymbol{q}} \left( a_{\boldsymbol{q}}a_{\boldsymbol{q}}^{\dagger} + \frac{1}{2} \right)\\ &+ \sum_{\boldsymbol{q}} g_{\boldsymbol{q}} \left( a_{\boldsymbol{q}} + a_{-\boldsymbol{q}}^{\dagger} \right)\exp \left( i\boldsymbol{q} \cdot \boldsymbol{r} \right)
\end{split}
\end{equation}
In the Hamiltonian above, we have $c_{\mathbf{p}}$ $(c_\mathbf{p}^{\dagger})$ as the creation (annihilation) operator for electrons with momentum $\mathbf{p}$ and energy $\varepsilon_{\mathbf{p}}$, whereas $a_{\mathbf{q}}$ $(a_\mathbf{q}^{\dagger})$ is the creation (annihilation) operator for longitudinal acoustic phonons of wave vector $\mathbf{q}$ and energy $\hbar \omega_{\mathbf{q}}$, respectively. Moreover, we assume a linear and isotropic dispersion $\omega_{\bm{q}} = v_s \vert \bm{q} \vert$, where $v_s$ is the speed of sound in the material. The electron-phonon interaction described to lowest-order (linear) is defined by its Fourier components as 
\begin{equation}
 g_{\bm q}=E_d \sqrt{\frac{\hbar}{2\rho\mathcal{V}v_s}q}
\end{equation}
where $\rho$ is mass density of the lattice, $\mathcal{V}$ is volume (area in two dimensions) and $E_d$ is the deformation potential constant characterizing the strength of the electron-phonon interaction arising from the modulation of the electronic energy due to lattice vibration.

Within the effective mass approximation, the single electron system part is simply
\begin{equation}\label{eq_app:System_Hamiltonian_Frohlich}
    \mathcal{H}_S = \frac{\vert \boldsymbol{p}\vert^2}{2m} =  \frac{\hbar^2 \vert \boldsymbol{k} \vert^2}{2m}
\end{equation}
where $m$ is taken as the renormalized mass due to the given band structure. The mean-field part arises as an expectation value of the lattice interaction over the multimode coherent state $\vert \boldsymbol{\alpha}\rangle$, yielding 
\begin{equation}\label{eq_app:deformation_potential}
\begin{split}
    \mathcal{H}_{\textrm{mf}} 
    &= \langle \boldsymbol{\alpha} \vert  \sum_{\mathbf{q}} g_{\mathbf{q}} \Big(a_{\mathbf{q}} + a_{\mathbf{-q}}^{\dagger} \Big) \vert \boldsymbol{\alpha} \rangle\\ &=
    \sum_{\substack{\mathbf{q}}}^{\vert \mathbf{q} \vert \le q_D}
    2g_{\mathbf{q}}
    \sqrt{\bar{n}_{\boldsymbol{q}}}
    \cos(\mathbf{q}\cdot\mathbf{r}-\omega_{\mathbf{q}}t+\varphi_{\mathbf{q}}).
\end{split}    
\end{equation}
This displacement field of lattice vibrations corresponds to the deformation potential (latter term) when determined with respect to a thermal quasi-classical coherent state
\begin{equation*}
    \alpha_\mathbf{q}=\sqrt{\bar{n}_{\boldsymbol{q}}}\exp(i\varphi_\mathbf{q}),
\end{equation*}
where $\bar{n}_{\boldsymbol{q}}$ is the thermal amplitude given by the Bose-Einstein distribution and $\varphi_{\mathbf{q}}=\arg(\alpha_{\mathbf{q}})$ is the random phase of the coherent state determining the initial conditions. We restrict the wave vector $\boldsymbol{q}$ to Debye wave number (isotropic cutoff) $q_D$ originating from the minimal lattice spacing $a$. The deformation potential similar to Eq.~\ref{eq_app:deformation_potential} has been investigated in previous studies~\cite{aydin_pnas_121_e2404853121_2024, KeskiRahkonen_phys.rev.lett_132_186303_2024, aydin_pnas_122_e2426518122_2025, Donghwan_phys.rev.b_106_054311_2022}

\subsection{Simulation parameters}

Here we present the full list of the simulation parameters employed to produce the results in the main text. Table~\ref{tabel: system parameters} contains the relevant material parameters, such as the effective mass, sound speed and lattice constant, for determining the dynamics within the deformation potential scheme. Other necessary ingredients, i.e., the Debye cutoff, temperature and Fermi momentum respectively are derived from these parameters as
\begin{align}
    \label{eq_app:Parameters}
    q_D= \frac{2\sqrt\pi}{a} 
    \quad \& \quad
    k_F= q_D/2
    \quad \& \quad
    T_D=\frac{q_D\hbar v_s}{k_B},
\end{align}
which are also listed in Table~\ref{tabel: system parameters}. These parameters are consistent with the previous studies~\cite{aydin_pnas_121_e2404853121_2024, KeskiRahkonen_phys.rev.lett_132_186303_2024, aydin_pnas_122_e2426518122_2025, Donghwan_phys.rev.b_106_054311_2022}.
We also point out the robustness of the results reported in the main text over a reasonable range of material parameters, in a similar manner as analyzed in Ref.~\cite{aydin_pnas_121_e2404853121_2024}. In other words, reasonable deviations from the given values do not change the qualitative conclusions of the work.

\begin{table}[h!]
\centering
\begin{tabular}{cccccc}
Material & $m$ & $v_s$ & a & $E_d $ & $\rho$    \\ 
&$[m_0]$& $[\textrm{m/s}]$ & $[\textrm{\AA}]$ & [\textrm{eV}] & $[\textrm{kg/m}^3]$\\
\hline
\\
Copper   & $\sim1$                    & 4700 & 0.36 & 10 & 8960 \\
\\
Bi2212   & 8.4                  & 2800 & 0.54 & 10 & 5200 \\\\ \hline
\end{tabular}
\caption{System parameters employed in the simulations regarding the Fr{\"o}hlich model.}
\label{tabel: system parameters}
\end{table}

The strange metal (Bi2212) in Tab.~\ref{tabel: system parameters} possess two characteristic attributes: a relatively high deformation potential constant and low Fermi energy compared to normal metals, such as copper. That is to say, we can divide the system dynamics into the two classes of weak and strong lattice interactions in the following way
\begin{align*}
\bar{K} = \frac{E_F}{\Delta V_\textrm{def}}
 \begin{cases}
	\gg 1  & \Rightarrow \; \textrm{perturbative} \Leftarrow \; \textrm{copper}, \\
    \\
	\le 1 \;  & \Rightarrow \;  \textrm{nonperturbative} \Leftarrow \; \textrm{Bi2212},
\end{cases}
\end{align*}
where the root-mean-square identifies the strength of lattice disorder $\Delta V_\textrm{def}$, growing in temperature as
\begin{equation*}
 \Delta V_\textrm{def}^2 = \frac{2 E_d^2 \hbar}{\pi \rho v_s} \int_0^{q_D} \frac{q^2\textrm{d} q}{e^{\hbar v_s q/k_BT}-1}.
\end{equation*}
In the present study, this type of classification holds for both prototypical metals across all temperatures considered.

\subsection{Time propagation}

Within the quantum-acoustical framework, the system dynamics is governed the master equation
\begin{equation}
\begin{split}
    i \frac{\partial}{\partial t}|\psi_{\pm}\rangle &=\Bigg[\mathcal{H}_{S}+ \mathcal{H}_{B+I}^{\pm}\Bigg]|\psi_{\pm}\rangle,
\end{split}   
\label{eq_app:time evolution}
\end{equation}
where we have defined
\begin{equation*}
    \mathcal{H}_{B+I}^{\pm}= \mathcal{H}_{\textrm{mf}} - \sum_{\bm q} \Big(\bm \eta_{\bm q}(t) \cdot \bm g_{\bm q} \mp \frac{1}{2}\bm \nu_{\bm q}(t)\bm \cdot \bm g_{\bm q} \Big).
\end{equation*}
This equation of motion was derived previously [see Eq.~\ref{eq_app:equation_of_motion_2}]. By comparing the interaction part of the Fr{\"o}hlich model in Eq.~\ref{Eq_app:Frohlich_Hamiltonian} with the nominal interaction Hamiltonian in Eq.~\ref{Eq_app:interaction_Hamiltonian}, we see that the coupling parameter for this model is
\begin{equation}
    \bm g_{\bm q}(\bm r)=
    \frac{E_d |\bm q|}{\sqrt{\rho\mathcal{V}\omega_{\bm q}}}
    \begin{bmatrix}
    \cos({\bm q\cdot\boldsymbol{r}} + \pi)\\
    \sin({\bm q\cdot\boldsymbol{r}})
    \end{bmatrix}.
\end{equation}
The corresponding system and the mean-field Hamiltonians are presented in Eqs.~\ref{eq_app:System_Hamiltonian_Frohlich} and \ref{eq_app:deformation_potential}, respectively. In addition, we need to define Gaussian-distributed noise variables $\boldsymbol{\eta}_{\boldsymbol{q}}$ and $\boldsymbol{\nu}_{\boldsymbol{q}}$. However, the mean-field Hamiltonian and the noise can be combined into a single stochastic pseudopotential entailing the lattice interaction on the system $\mathcal{H}_{B+I}^{\pm}$ whose generation will be discussed after we outline the numerical scheme for solving the master equation. 

The quantum-acoustical master equation above admits a formal solution that can be leveraged in developing a split-operator strategy in the following way
\begin{equation}
\begin{split}
    |\psi_{\pm}(t)\rangle &=\exp\left({-i\int_0^t ds \mathcal{H}_{S}+ \mathcal{H}_{B+I}^{\pm} (\bm r,s)}\right)|\psi_{\pm}(0)\rangle\\
    &=\exp\left({-i\Delta t\sum_j  \mathcal{H}_{S}+ \mathcal{H}_{B+I}^{\pm} (\bm r,j\Delta t)}\right)|\psi_{\pm}(0)\rangle\\&+\mathcal{O}(\Delta t)\\
    &= \prod_j \Big[\exp(-i \mathcal{H}_{B+I}^{\pm}(j\Delta t ) \Delta t/2)\exp(-i\mathcal{H}_{S}\Delta t)\\& \times\exp(-i \mathcal{H}_{B+I}^{\pm} (j\Delta t ) \Delta t/2) \Big] |\psi_{\pm}(0)\rangle\\ &+\mathcal{O}(\Delta t)+\mathcal{O}(\Delta t^2 \eta).
\end{split}
\end{equation}
The leading order $\mathcal{O}(\Delta t)$ error stems from the approximation of the integral by a Riemann sum, whereas the $\mathcal{O}(\Delta t^2 \eta)$ error follows from the non-commutativity of the Hamiltonian between two different timesteps within the approximation based on the Baker-Campbell-Hausdorff formula. Here, the variable $\eta$ in the error is proportional to the noise. However, by employing the midpoint rule for the Riemann sum, the integration error can be reduced one level down to $\mathcal{O}(\Delta t^2)$. Interestingly, for the quantum-acoustical mean-field potential, the difference between the midpoint rule and the left rule, which is just a shift in time, only amounts to a shift of the initial phases of each coherent mode. As these are random, the midpoint and left rules are equivalent in this case, leading to an error of $\mathcal{O}(\Delta t^2)$ instead of $\mathcal{O}(\Delta t)$.

In order to utilize the master equation, we use the following efficient scheme to generate the stochastic pseudopotential $\mathcal{H}_{B+I}^{\pm}$ containing noise terms $\bm \eta(t)$ and $\bm \nu(t)$ that are normally distributed with zero mean and covariance specified (Eq.~\ref{eq_app:noise_covariance_original} restated) as
\begin{equation} \label{eq_app:noise_covariance}
\begin{split}
\Big \langle \bm \eta^T_{\bm q}(s) \cdot \bm \eta_{\boldsymbol{r}}(t')\Big \rangle_W &=\frac{1}{2} \bar{\boldsymbol{L}}_{\bm q}(t'-s)\delta_{\bm q,\boldsymbol{r}},
\\
\\
\Big \langle \bm\eta^T_{\boldsymbol{q}}(s)\bm \cdot \bm \nu_{\bm r}(t')\Big \rangle_W &=i\bar{\boldsymbol{M}}_{\bm q}(t'-s)\Theta(s-t')\delta_{\bm q,\boldsymbol{r}},
\\
\\
\Big \langle \bm\nu^T_{\bm q}(s) \cdot \bm\nu_{\boldsymbol{r}}(t')\Big \rangle_W &=0.
\end{split}
\end{equation}
This covariance is, however, not positive semi-definite, thus yielding an ill-defined normal distribution. As a remedy to this complication, we let the noise variables be complex instead of real-valued. Consequently, this extension leads to extra covariances that are not restricted physically, but we take advantage of them to freely choose the covariances:
\begin{equation}
        \Big \langle \bm \eta^T_{\bm q}(s) \cdot \overline{\bm{\eta}}_{\boldsymbol{r}}(t')\Big \rangle_W, 
        \Big \langle \bm \eta^T_{\bm q}(s) \cdot \overline{\bm{\nu}}_{\boldsymbol{r}}(t')\Big \rangle_W, 
        \Big \langle \bm \nu^T_{\bm q}(s) \cdot \overline{\bm{\nu}}_{\boldsymbol{r}}(t')\Big \rangle_W.
\end{equation}
In particular, we can construct these extra, unphysical covariances in such a way that the overall covariance becomes positive semi-definite~\cite{stockburger_chem.phys.296.159.2004}. 

Determining the appropriate form of these covariances can be highly nontrivial. A convenient and efficient method for generating the noise is to employ the fast Fourier transform $\mathcal{F}$, which inherently enforces the desired covariances~\cite{stockburger_chem.phys.296.159.2004}. In this two-step strategy, we define the covariance matrix with the property of Eq.~\ref{eq_app:noise_covariance} as
\begin{equation}
    \text{Cov}(\bm \eta_{\bm q},\bm \nu_{\bm q})(t)=
    \begin{bmatrix}
         \Big \langle \bm \eta^T_{\bm q}(0) \cdot \bm{\eta}_{\boldsymbol{r}}(t)\Big \rangle_W & \Big \langle \bm\eta^T_{\boldsymbol{q}}(0)\bm \cdot \bm \nu_{\bm r}(t)\Big \rangle_W\\
         \Big \langle \bm\eta^T_{\boldsymbol{q}}(t)\bm \cdot \bm \nu_{\bm r}(0)\Big \rangle_W & \Big \langle \bm\nu^T_{\bm q}(0) \cdot \bm\nu_{\boldsymbol{r}}(t)\Big \rangle_W
    \end{bmatrix},
\end{equation}
and we generate normally distributed white noise $\bm w_{\bm q}(t)$ satisfying
\begin{equation}
    \langle \bm w_{\bm q}(s)\rangle=0
    \quad \textrm{and} \quad
    \langle\bm  w_{\bm q}(s)\bm w_{\bm q}(t')\rangle=\delta(t'-s).
\end{equation}
This noise $\bm w_{\bm q}(t)$ should have 4 components for each $\bm q$ and timestep, i.e. the same size as a row of the covariance matrix above. Next, the noise with the correct covariance property is then given by
\begin{equation}
    \begin{bmatrix}
        \bm \eta_{\bm q}\\
        \bm \nu_{\bm q}
    \end{bmatrix}
    =\mathcal{F}^{-1}\left\{\sqrt{\mathcal{F}\left[\text{Cov}(\bm \eta_{\bm q},\bm \nu_{\bm q})\right]}\mathcal{F}\left[\bm w_{\bm q}\right]\right\},
\end{equation}
where we have used the matrix root.

\subsection{Initial state}

To investigate the dynamics governed by the quantum-acoustical master equation defined in Eq.~\ref{eq_app:time evolution}, we must first specify an initial state of the system $\mathcal{H}_s$. We approach this matter by considering Gaussian wavepackets, a common and effective tool for analyzing the time-dependent behavior of quantum systems~\cite{heller2018semiclassical, tannor2007introduction}, as employed, e.g., in studies of quantum optics~\cite{scully1997quantum, walls2007quantum}, scarring~\cite{keski2019quantum, keskirahkonen_PhysRevB.96.094204.2017, keski2024antiscarring}, and branched flow~\cite{superwire1, superwire2}, as well as for analyzing deformation potential dynamics~\cite{aydin_pnas_121_e2404853121_2024, aydin_pnas_122_e2426518122_2025, Donghwan_phys.rev.b_106_054311_2022, zimmermann_entropy_26_552_2024}. Specifically, we select the following test Gaussian to represent the initial state of the charge carrier:
\begin{equation}\label{Eq_app:Initial_Gaussian}
    \psi(\mathbf{r}, 0) = \mathcal{N}\exp\left( \frac{1}{4} \vert \mathbf{r} \cdot \boldsymbol{\sigma} \vert^2  - i \mathbf{k} \cdot \mathbf{r} \right),
\end{equation}
where $\mathcal{N}$ is the normalization factor, with $\boldsymbol{\sigma} = (\sigma_x^{-1}, \sigma_y^{-1})$ describing the initial width of the wavepacket. To ensure the charge carrier begins at a well-defined position, away from the system boundaries, we set $\sigma = 0.02L$, with $L$ denoting the physical size of the simulated space. Without loss of generality, we choose to launch the wavepacket along the $x$-direction with the Fermi momentum, such that $\mathbf{k} = (k_F, 0)$, where $k_F$ is the Fermi wavevector. 

The chosen Gaussian wavepacket emulates a charge carrier injected into the material at the Fermi surface.
However, for the studied range of temperatures, the memory of the initial form of the wavepacket is quickly lost in the chaotic interaction with the lattice, and its exact form hence becomes irrelevant. 

\subsection{Relaxation time and wavepacket spread}

Given that the time evolution of the states $|\psi_{\pm}\rangle$ is governed by the master equation in Eq.~\ref{eq_app:time evolution}, the instantaneous expectation value of any observable associated with the Hermitian operator $\hat{O}$ can be computed accordingly
\begin{equation}\label{eq_app:expectation_value}
    \langle\hat O \rangle (t) =\left<\langle\psi_-(t)|\hat O|\psi_+(t) \rangle\right>_W.
\end{equation}

For instance, this quantum-acoustical approach enable us to study the momentum relaxation in the direction of the launched wavepacket under the influence of lattice vibrations. Within the relaxation time approximation, the magnitude of the average momentum is expected to decay as
\begin{equation}
\label{eq_app:momentum relax}
    p_x(t)=p_x(0)e^{-t/\tau}
\end{equation}
that allows us to estimate the inverse momentum relaxation time $1/\tau$ by fitting the average momentum to the exponential form.

In the weak coupling regime ($\bar{K} \gg 1$), the exponential decay of the average momentum of the wave packet is present. Furthermore, Fermi’s golden rule works very well within this domain~\cite{Donghwan_phys.rev.b_106_054311_2022}. In particular, the numerical result can then be compared with the perturbative result of Sec.~\ref{sec: Perturbation theory}. However, when the wavelength of the charge carrier is larger than the shortest length scale of the mean-field, the average momentum can show nonexponential decay due to quantum coherence and interference effects~\cite{Donghwan_phys.rev.b_106_054311_2022}. Nevertheless, the extracted $\tau$ value still gives a qualitative description for the scattering rate. 

In contrast, the coupling between the charge carrier and lattice vibrations becomes highly nonperturbative when the parameter $\bar{K} \le 1$. Within this strong interaction regime, charge carriers scatter off lattice vibrations, but the interaction can also lead to short-time localization events~\cite{Donghwan_phys.rev.b_106_054311_2022, zimmermann_entropy_26_552_2024, KeskiRahkonen_phys.rev.lett_132_186303_2024, aydin_pnas_121_e2404853121_2024}. Even under this type of transient localization, the momentum of the wavepacket generally decays over time, albeit in a highly nonmonotonic manner, rendering the scattering rate $\tau$ extracted from the exponential fit as an unreliable measure. Moreover, no analytical results are available to serve as a benchmark for this regime.

To investigate the general dynamics governed by the quantum-acoustical master equation in a manner that is independent of the coupling strength $\bar{K}$, we consider the time-averaged spatial spread of the wavepacket, defined as
\begin{equation}
\begin{split}
    \xi &= \sqrt{\frac{1}{T} \int_0^T  \left<\langle\psi_-|\Delta x^2+\Delta y^2|\psi_+\rangle\right>_W \, dt}\\
    &= \Bigg( \frac{1}{T} \int_0^T  \left<\langle\psi_-|x^2+y^2|\psi_+\rangle\right>_W\\&-\left<\langle\psi_-|x|\psi_+\rangle\right>_W^2-\left<\langle\psi_-|y|\psi_+\rangle\right>_W^2 \, dt \Bigg)^{1/2},
\end{split}
\end{equation}
where the averaging time window is set to $T = 40\, \textrm{fs}$. It provides an overall sense of how spread out the wavepacket is over a chosen time window. 
In systems exhibiting localization, the wavepacket eventually ceases to spread further, allowing $\xi$ in the limit of $T \rightarrow \infty$ to serve as a proxy for the localization length. However, in our setting, localization effects do not emerge within the finite time window, but even with longer simulations, any apparent localization would be transient due to the evolving lattice disorder (see, e.g., Refs.~\cite{Donghwan_phys.rev.b_106_054311_2022, zimmermann_entropy_26_552_2024, KeskiRahkonen_phys.rev.lett_132_186303_2024, aydin_pnas_121_e2404853121_2024}). Nonetheless, the defined measure $\xi$ remains sensitive to the nature of the wavepacket dynamics and serves as a simple, effective indicator to distinguish between ballistic, diffusive, and localized regimes.


\subsection{Perturbation theory}
\label{sec: Perturbation theory}

Here we evaluate the transport scattering rate according to the perturbation theory within the coherent-state picture that is employed as a benchmark for the master equation simulations in the weak coupling limit ($\bar{K} \gg 1$). Following the steps established in Ref.~\cite{Donghwan_phys.rev.b_106_054311_2022}, we consider the lattice with a multimode coherent
state $\vert \boldsymbol{ \alpha} \rangle = \otimes_{\boldsymbol{q}} \vert \alpha_{\boldsymbol{q}} \rangle$ and then study the scattering of an initial many-body state $\vert \boldsymbol{k}, \boldsymbol{\alpha} \rangle$, where $\boldsymbol{k}$ is an electron wavevector due to the lattice-electron interaction. The inverse of the momentum relaxation time $\tau$ can be determined from the momentum autocorrelation within the interaction picture. For the Fr{\"o}hlich model, it is given by
\begin{equation}
\begin{split}
    \frac{1}{\tau} =&-\sum_{\bm q}\frac{\bm k\cdot\bm q}{\bm k^2}\frac{2\pi g_{\bm q}^2}{\hbar}\Big[N_{\bm q}\delta(\epsilon(\bm k +\bm q)-\epsilon(\bm k)-\hbar\omega_{\bm q})\\ &+(N_{\bm q}+1)\delta(\epsilon(\bm k +\bm q)-\epsilon(\bm k)+\hbar\omega_{\bm q})\Big].
\end{split}    
\end{equation}

Subsequently, we introduce the proper Fermi statistics in the scattering rate as
\begin{align*}
    \frac{1}{\tau} =&-\beta \int_0^{\infty}d\epsilon(\bm k)\sum_{\bm q}\frac{\bm k\cdot\bm q}{\bm k^2}\frac{2\pi g_{\bm q}^2}{\hbar}\\&\times\Big[
    N_{\bm q}\delta(\epsilon(\bm k +\bm q)-\epsilon(\bm k)-\hbar\omega_{\bm q})\\&\times f(\epsilon(\bm k))\{1-f(\epsilon(\bm k)+\hbar\omega_{\bm q})\}\\&+(N_{\bm q}+1)\delta(\epsilon(\bm k +\bm q)-\epsilon(\bm k)+\hbar\omega_{\bm q})\\&\times f(\epsilon(\bm k))\{1-f(\epsilon(\bm k)-\hbar\omega_{\bm q})\}]\Big],
\end{align*}
where $\beta = 1/(k_{\textrm{B}}T)$ is the inverse temperature and $f(\epsilon) = [1 + \exp(-\beta \epsilon)]^{-1}$ is the Fermi function. Physically, we can interpret that $f(\epsilon(\bm k))$ represents the probability for the initial state with energy $\epsilon(\bm k)$ to be occupied, and vice versa $1-f(\epsilon(\bm k)\pm\hbar\omega_{\bm q})$ the probability for the final state with energy $\epsilon(\bm k)\pm\hbar\omega_{\bm q}$ to be unoccupied. 

Next, utilizing the density-of-state approximation to transform the summation into an integration over the $k$ space, we estimate the scattering rate to be
\begin{align}
\begin{split}
    \frac{1}{\tau}  &\approx -\beta \frac{A}{(2\pi)^2} \int_0^{\infty}dk \frac{\hbar^2 k}{m^*} \int_0^{2\pi}d\phi\int_0^{q_D}dq q\\ &\frac{ kq\cos{\phi}}{\bm k^2}\frac{2\pi g_{\bm q}^2}{\hbar} \Bigg[
    N_{\bm q}\delta(\frac{\hbar^2 kq\cos{\phi}}{m^*}+\frac{\hbar^2 q^2}{2m^*}-\hbar\omega_{\bm q})\\ &\times f(\epsilon(\bm k))\{1-f(\epsilon(\bm k)+\hbar\omega_{\bm q})\}\\ &+(N_{\bm q}+1)\delta(\frac{\hbar^2 kq\cos{\phi}}{m^*}+\frac{\hbar^2 q^2}{2m^*}+\hbar\omega_{\bm q})\\ &\times f(\epsilon(\bm k))\{1-f(\epsilon(\bm k)-\hbar\omega_{\bm q})\}\Bigg],
\end{split}
\end{align}
where we have estimated the density of states as that of a two-dimensional free-electron gas, i.e., the density of states is given by $A/(2\pi)^2$ with $A$ the area of the system. Since the argument of the delta function vanishes when
\begin{equation}\label{eq_app:delta arg zero}
\begin{split}
    \frac{\hbar^2kq\cos(\phi)}{m^*}+\frac{\hbar^2 q^2}{2m^*}\pm \hbar\omega_{\bm q} = 0,
\end{split}
\end{equation}
subsequently meaning
\begin{equation}
  \cos(\phi)=-\frac{q}{2k}\mp\frac{m^* v_s}{\hbar k}.  
\end{equation}
we have the condition
\begin{equation}
\label{eq_app:k_condition}
    |\frac{q}{2}\pm\frac{m^* v_s}{\hbar }|\leq k.
\end{equation}
Notably, there are two values $\phi$ such that the condition in Eq.~\ref{eq_app:delta arg zero} holds, so we multiply by an additional factor of 2. Furthermore, we can perform the angle integration by noting that the derivative of the argument of the delta function is $\frac{\hbar^2kq\sin(\phi)}{m^*}$ and when the argument of the delta function vanishes this derivative equals
\begin{equation}
    \frac{\hbar^2kq\sqrt{1-(\frac{q}{2k}\pm\frac{m^* v_s}{\hbar k})^2}}{m^*}.
\end{equation}

Finally, we end up with the following estimation of the scattering rate
\begin{widetext}
\begin{align}
\begin{split}
    \frac{1}{\tau} =&\beta \frac{E_d^2 A}{2\pi\rho \mathcal{V}v_s}  \int_0^{q_D}dq \int_{|q/2-m^*v_s/\hbar|}^{\infty}dk \frac{ q^2(\frac{q}{2}-\frac{ m^*v_s}{\hbar })}{k^2\sqrt{1-(\frac{q}{2k}-\frac{m^* v_s}{\hbar k})^2}} 
    N_{\bm q}f(\epsilon(\bm k))\{1-f(\epsilon(\bm k)+\hbar\omega_{\bm q})\}\\
    &+\beta \frac{E_d^2 A}{2\pi\rho\mathcal{V}v_s} \int_0^{q_D}dq \int_{q/2+m^*v_s/\hbar}^{\infty}dk \frac{ q^2(\frac{q}{2}+\frac{m^* v_s}{\hbar })}{ k^2\sqrt{1-(\frac{q}{2k}+\frac{m^* v_s}{\hbar k})^2}} (N_{\bm q}+1)f(\epsilon(\bm k))\{1-f(\epsilon(\bm k)-\hbar\omega_{\bm q})\}
\end{split}
\end{align}
\end{widetext}
that has been employed in the main text as a point of reference for the master-equation simulations in the nonperturbative regime. Physically, the scattering rate above is composed of two factors: The former term describes the absorption caused by the lattice vibrations; whereas the latter term characterizes both the stimulated and spontaneous emission. Basically the derived scattering time formula is a slightly more general version of the previous works of Hwang and Das Sarma~\cite{Hwang_phys.rev.b_77_115449_2008} and Efetov and Kim~\cite{Kim_phys.rev.lett_105_256805_2010}. As analyzed in Ref.~\cite{Donghwan_phys.rev.b_106_054311_2022}, this result is also consistent with the conventional Bloch-Gr{\"u}neisen theory~\cite{ziman2001electrons}.

On the other hand, we can also analyze the scattering rate with respect to the mean-field approach where one only considers the scattering stemming from the deformation potential defined in Eq.~\ref{eq_app:deformation_potential} according to the thermal multimode coherent state $\vert \boldsymbol{\alpha} \rangle$. By carrying out the same machinery as for the full treatment, the perturbation theory for the mean-field method yields the following scattering rate:
\begin{widetext}
\begin{align}
\begin{split}
    \frac{1}{\tau} =&\beta \frac{E_d^2 A}{2\pi\rho \mathcal{V}v_s}  \int_0^{q_D}dq \int_{|q/2-m^*v_s/\hbar|}^{\infty}dk \frac{ q^2(\frac{q}{2}-\frac{ m^*v_s}{\hbar })}{k^2\sqrt{1-(\frac{q}{2k}-\frac{m^* v_s}{\hbar k})^2}} 
    N_{\bm q}f(\epsilon(\bm k))\{1-f(\epsilon(\bm k)+\hbar\omega_{\bm q})\}\\
    &+\beta \frac{E_d^2 A}{2\pi\rho\mathcal{V}v_s} \int_0^{q_D}dq \int_{q/2+m^*v_s/\hbar}^{\infty}dk \frac{ q^2(\frac{q}{2}+\frac{m^* v_s}{\hbar })}{ k^2\sqrt{1-(\frac{q}{2k}+\frac{m^* v_s}{\hbar k})^2}} N_{\bm q}f(\epsilon(\bm k))\{1-f(\epsilon(\bm k)-\hbar\omega_{\bm q})\}.
\end{split}
\end{align}
\end{widetext}
that is consistent with the previous work~\cite{Donghwan_phys.rev.b_106_054311_2022} on the deformation potential scattering. Notably, the scattering rate only deviates from the full treatment by missing spontaneous emission, i.e., the term $N_{\bm q}+1$ is instead replaced by $N_{\bm q}$. This result is natural since the spontaneous emission is associated with the absent quantum fluctuations in the mean-field. Since spontaneous emission is absent in a deformation field, detailed balance is broken and a charge carrier within the deformation field will heat up as time goes by. In the
high temperature limit, where all the mode energies are small compared to the thermal energy, the difference between the full quantum and mean-field approaches within perturbation theory vanishes as detailed in Ref.~\cite{Donghwan_phys.rev.b_106_054311_2022}. 

\bibliography{apssamp}

@PREAMBLE{
 "\providecommand{\noopsort}[1]{}" 
 # "\providecommand{\singleletter}[1]{#1}%" 
}

@article{cea_phys.rev.b_100_205113_2019,
    title={{Electronic band structure and pinning of Fermi energy to Van Hove singularities in twisted bilayer graphene: A self-consistent approach}},
    author={Cea, Tommaso and Walet, Niels R and Guinea, Francisco},
    journal={Phys. Rev. B},
    volume={100},
    number={20},
    pages={205113},
    year={2019},
    publisher={APS},
    doi = {10.1103/PhysRevB.100.205113}
}

@article{lin_phys.rev.x_3_021002_2013,
    title={Fermi surface of the most dilute superconductor},
    author={Lin, Xiao and Zhu, Zengwei and Fauqu{\'e}, Beno{\^\i}t and Behnia, Kamran},
    journal={Phys. Rev. X},
    volume={3},
    number={2},
    pages={021002},
    year={2013},
    publisher={APS},
    doi = {10.1103/PhysRevX.3.021002}
}

@article{zhang_nat.commun_15_3737_2024,
    title={Observation of dichotomic field-tunable electronic structure in twisted monolayer-bilayer graphene},
    author={Zhang, Hongyun and Li, Qian and Park, Youngju and Jia, Yujin and Chen, Wanying and Li, Jiaheng and Liu, Qinxin and Bao, Changhua and Leconte, Nicolas and Zhou, Shaohua and others},
    journal={Nat. Commun.},
    volume={15},
    number={1},
    pages={3737},
    year={2024},
    publisher={Nature Publishing Group UK London},
    doi = {10.1038/s41467-024-48166-8}
}

@article{nimbalkar_nanomicro.letter_12_1_2020,
    title={Opportunities and challenges in twisted bilayer graphene: a review},
    author={Nimbalkar, Amol and Kim, Hyunmin},
    journal = {Nano-Micro Lett.},
    fjournal={Nano-Micro Letters},
    volume={12},
    pages={1--20},
    year={2020},
    publisher={Springer},
    doi = {10.1007/s40820-020-00464-8},
}

@article{skipetrov_j.alloys.compd_893_162330_2022,
    title={Electronic structure and unusual magnetic properties of diluted magnetic semiconductors $\textrm{Pb}_{1-x-y}\textrm{Sn}_x\textrm{Sc}_y\textrm{Te}$},
    author={Skipetrov, EP and Bogdanov, EV and Kovalev, BB and Skipetrova, LA and Knotko, AV and Emelyanov, AV and Taldenkov, AN and Slynko, VE},
    journal = {J. Alloys Compd.},
    fjournal={Journal of Alloys and Compounds},
    volume={893},
    pages={162330},
    year={2022},
    publisher={Elsevier},
    doi = {10.1016/j.jallcom.2021.162330}
}

@article{lafalce_acs.nano_18_18299_2024,
    title={{Optical studies of doped two-dimensional lead halide perovskites: Evidence for Rashba-split branches in the conduction band}},
    author={Lafalce, Evan and Bodin, Rikard and Larson, Bryon W and Hao, Ji and Haque, Md Azimul and Huynh, Uyen and Blackburn, Jeffrey L and Vardeny, Zeev Valy},
    journal={ACS nano},
    volume={18},
    number={28},
    pages={18299--18306},
    year={2024},
    publisher={ACS Publications},
    doi = {10.1021/acsnano.4c01525}
}

@article{han_energy.mater_5_0116_2024,
    title={Chalcogenide Perovskite $\textrm{Eu}\textrm{Hf}\textrm{S}_3$ with Low Band Gap and Antiferromagnetic Properties for Photovoltaics},
    author={Han, Yanbing and Fang, Jiao and Zhang, Han and Sun, Yiyang and Yuan, Yifang and Chen, Xu and Jia, Mochen and Li, Xinjian and Gao, Han and Shi, Zhifeng},
    journal = {Energy Mater.},
    fjournal={Energy Material Advances},
    volume={5},
    pages={0116},
    year={2024},
    publisher={AAAS},
    doi = {10.34133/energymatadv.0116},
}

@article{foulkes_rev.mod.phys_73_33_2001,
    title = {Quantum Monte Carlo simulations of solids},
    author = {Foulkes, W. M. C. and Mitas, L. and Needs, R. J. and Rajagopal, G.},
    journal = {Rev. Mod. Phys.},
    volume = {73},
    issue = {1},
    pages = {33--83},
    numpages = {0},
    year = {2001},
    month = {Jan},
    publisher = {American Physical Society},
    doi = {10.1103/RevModPhys.73.33},
    url = {https://link.aps.org/doi/10.1103/RevModPhys.73.33}
}

@article{ehrenfest_z.phys_45_455_1927,
    author={Ehrenfest, P.},
    title={{Bemerkung {\"u}ber die angen{\"a}herte G{\"u}ltigkeit der klassischen Mechanik innerhalb der Quantenmechanik}},
    journal = {Z. Phys.},
    fjournal={Zeitschrift f{\"u}r Physik},
    year={1927},
    month={Jul},
    day={01},
    volume={45},
    number={7},
    pages={455-457},
    issn={0044-3328},
    doi={10.1007/BF01329203},
    url={https://doi.org/10.1007/BF01329203}
}

@article{tanimura_j.phys.soc.jpn_58_101_1989, 
    author = {Tanimura, Yoshitaka and Kubo, Ryogo},
    title = {{Time evolution of a quantum system in contact with a nearly Gaussian-Markoffian noise bath}},
    journal = {J. Phys. Soc. Jpn.},
    fjournal = {Journal of the Physical Society of Japan},
    volume = {58},
    number = {1},
    pages = {101-114},
    year = {1989},
    doi = {10.1143/JPSJ.58.101},
    URL = {https://doi.org/10.1143/JPSJ.58.101}
}

@article{mccutcheon_phys.rev.b_84_081305_2011,
    author = {{McCutcheon}, Dara P.~S. and {Dattani}, Nikesh S. and {Gauger}, Erik M. and {Lovett}, Brendon W. and {Nazir}, Ahsan},
    title = {A general approach to quantum dynamics using a variational master equation: Application to phonon-damped Rabi rotations in quantum dots},
    journal = {Phys. Rev. B},
    year = {2011},
    month = aug,
    volume = {84},
    number = {8},
    pages = {081305},
    doi = {10.1103/PhysRevB.84.081305}
}

@article{hussey_philos.mag._84_2847_2004,
    issn = {1478-6435},
    journal = {Philos. Mag.},
    fjournal = {Philosophical Magazine},
    pages = {2847--2864},
    volume = {84},
    publisher = {Taylor & Francis Group},
    number = {27},
    year = {2004},
    title = {{Universality of the Mott-Ioffe-Regel limit in metals}},
    copyright = {Copyright Taylor & Francis Group, LLC 2004},
    address = {Abingdon},
    author = {Hussey, N. E. and Takenaka, K. and Takagi, H.},
    doi = {10.1080/14786430410001716944}
}

@article{gauber_phys_rev_131_2766_1963,
    title = {Coherent and Incoherent States of the Radiation Field},
    author = {Glauber, Roy J.},
    journal = {Phys. Rev.},
    volume = {131},
    issue = {6},
    pages = {2766--2788},
    numpages = {0},
    year = {1963},
    month = {Sep},
    publisher = {American Physical Society},
    doi = {10.1103/PhysRev.131.2766},
    url = {https://link.aps.org/doi/10.1103/PhysRev.131.2766}
}

@article{gunnarsson_rev.mod.phys_75_1085_2003,
    title = {Colloquium: Saturation of electrical resistivity},
    author = {Gunnarsson, O. and Calandra, M. and Han, J. E.},
    journal = {Rev. Mod. Phys.},
    volume = {75},
    issue = {4},
    pages = {1085--1099},
    numpages = {0},
    year = {2003},
    month = {Oct},
    publisher = {American Physical Society},
    doi = {10.1103/RevModPhys.75.1085},
    url = {https://link.aps.org/doi/10.1103/RevModPhys.75.1085}
}

@book{Weiss_book,
    author = {Weiss, Ulrich},
    title = {Quantum Dissipative Systems},
    publisher = {WORLD SCIENTIFIC},
    year = {2012},
    doi = {10.1142/8334},
    edition  = {4th},
    address={Singapore} 
}

@article{lee_phys.rev_90_297_1953,
    title = {The Motion of Slow Electrons in a Polar Crystal},
    author = {Lee, T. D. and Low, F. E. and Pines, D.},
    journal = {Phys. Rev.},
    volume = {90},
    issue = {2},
    pages = {297--302},
    numpages = {0},
    year = {1953},
    month = {Apr},
    publisher = {American Physical Society},
    doi = {10.1103/PhysRev.90.297},
    url = {https://link.aps.org/doi/10.1103/PhysRev.90.297}
}

@article{grusdt_phys.rev.a_90_063610_2014,
    title = {Bloch oscillations of bosonic lattice polarons},
    author = {Grusdt, F. and Shashi, A. and Abanin, D. and Demler, E.},
    journal = {Phys. Rev. A},
    volume = {90},
    issue = {6},
    pages = {063610},
    numpages = {23},
    year = {2014},
    month = {Dec},
    publisher = {American Physical Society},
    doi = {10.1103/PhysRevA.90.063610},
    url = {https://link.aps.org/doi/10.1103/PhysRevA.90.063610}
}

@article{de.vega_rev.mod.phys_89_015001_2017,
    title = {{Dynamics of non-Markovian open quantum systems}},
    author = {de Vega, In\'es and Alonso, Daniel},
    journal = {Rev. Mod. Phys.},
    volume = {89},
    issue = {1},
    pages = {015001},
    numpages = {58},
    year = {2017},
    month = {Jan},
    publisher = {American Physical Society},
    doi = {10.1103/RevModPhys.89.015001},
    url = {https://link.aps.org/doi/10.1103/RevModPhys.89.015001}
}

@article{Schmitz_eur.phys.j_227_1929_2019,
    journal = {Eur. Phys. J.},
    fjournal = {The European physical journal. ST, Special topics},
    pages = {1929--1937},
    volume = {227},
    publisher = {Springer Berlin Heidelberg},
    number = {15-16},
    year = {2019},
    title = {{A variance reduction technique for the stochastic Liouville–von Neumann equation}},
    copyright = {EDP Sciences, Springer-Verlag GmbH Germany, part of Springer Nature 2019},
    address = {Berlin/Heidelberg},
    author = {Schmitz, Konstantin and Stockburger, Jürgen T.},
    doi = {10.1140/epjst/e2018-800094-y}
}

@article{imai_chem.phys_446_134_2015,
    title = {{Application of stochastic Liouville–von Neumann equation to electronic energy transfer in FMO complex}},
    journal = {Chem. Phys.},
    fjournal = {Chemical Physics},
    volume = {446},
    pages = {134-141},
    year = {2015},
    doi = {https://doi.org/10.1016/j.chemphys.2014.11.014},
    url = {https://www.sciencedirect.com/science/article/pii/S0301010414003231},
    author = {Hajime Imai and Yukiyoshi Ohtsuki and Hirohiko Kono}
}

@article{dunn_j.chem.phys_150_184109_2019,
    author = {Dunn, Ian S. and Tempelaar, Roel and Reichman, David R.},
    title = {Removing instabilities in the hierarchical equations of motion: Exact and approximate projection approaches},
    journal = {J. Chem. Phys.},
    volume = {150},
    number = {18},
    pages = {184109},
    year = {2019},
    month = {05},
    issn = {0021-9606},
    doi = {10.1063/1.5092616},
    url = {https://doi.org/10.1063/1.5092616}
}

@article{kundu_annu.rev.phys.chem_73_349_2022,
   author = {Kundu, Sohang and Makri, Nancy},
   title = {Intramolecular Vibrations in Excitation Energy Transfer: Insights from Real-Time Path Integral Calculations}, 
   journal= {Annu. Rev. Phys. Chem.},
   year = {2022},
   volume = {73},
   pages = {349-375},
   doi = {https://doi.org/10.1146/annurev-physchem-090419-120202},
   url = {https://www.annualreviews.org/content/journals/10.1146/annurev-physchem-090419-120202},
   publisher = {Annual Reviews},
   issn = {1545-1593},
  }

@article{stockburger_europhys.lett_115_40010_2016,
    title={Exact propagation of open quantum systems in a system-reservoir context},
    author={Stockburger, J{\"u}rgen T},
    journal={Europhys. Lett.},
    volume={115},
    number={4},
    pages={40010},
    year={2016},
    publisher={IOP Publishing},
    doi = {10.1209/0295-5075/115/40010}
}

@book{Mahan_book,
  title={Many-particle physics},
  author={Mahan, Gerald D},
  year={2000},
  publisher={Springer Science \& Business Media},
  address={New York, NY, USA}
}

@book{kittel2018introduction,
    title={Introduction to Solid State Physics},
    author={Kittel, C.},
    isbn={9788126578436},
    year={2018},
    publisher={John Wiley \& Sons},
    address={Hoboken, NJ, USA}
}

@book{ashcroft1976solid,
    author={N.W. Ashcroft and N.D. Mermin},
    title={Solid State Physics},
    isbn={9780030839931},
    lccn={lc74009772},
    series={HRW international editions},
    year={1976},
    publisher={Holt, Rinehart and Winston},
    address={New York, NY, USA}
}

@book{walls2007quantum,
    title={Quantum Optics},
    author={Walls, D.F. and Milburn, G.J.},
    isbn={9783540285748},
    lccn={2007936291},
    year={2007},
    publisher={Springer},
    address={Berlin Heidelberg, Germany}
}

@book{scully1997quantum,
    title={Quantum Optics},
    author={Scully, M.O. and Zubairy, M.S.},
    isbn={9780521435956},
    lccn={94042949},
    year={1997},
    publisher={Cambridge University Press},
    address={Cambridge, UK}
}

@book{feynman1965quantum,
    title={Quantum Mechanics and Path Integrals},
    author={Feynman, R.P. and Hibbs, A.R.},
    isbn={9780070206502},
    lccn={64025171},
    series={International Earth and Planetary Sciences Series},
    year={1965},
    publisher={McGraw-Hill},
    address={New York, NY, USA}
}

@book{schulman_book,
  title={Techniques and applications of path integration},
  author={Schulman, Lawrence S},
  year={2012},
  publisher={Courier Corporation},
  address={Mineola, NY, USA}
}

@book{Altland_book,
  title={Condensed matter field theory},
  author={Altland, Alexander and Simons, Ben D},
  year={2010},
  publisher={Cambridge university press},
  address={Cambridge, UK}
}

@article{Giustino_rev.mod.phys_89_015003_2018,
    title = {Electron-phonon interactions from first principles},
    author = {Giustino, Feliciano},
    journal = {Rev. Mod. Phys.},
    volume = {89},
    issue = {1},
    pages = {015003},
    numpages = {63},
    year = {2017},
    month = {Feb},
    publisher = {American Physical Society},
    doi = {10.1103/RevModPhys.89.015003},
    url = {https://link.aps.org/doi/10.1103/RevModPhys.89.015003}
}

@article{Bloch_z.phys_59_208_1930,
    author    = {F. Bloch},
    title     = {{Zum elektrischen Widerstandsgesetz bei tiefen Temperaturen}},
    journal = {Z. Phys},
    fjournal = {Zeitschrift für Physik},
    year      = {1930},
    month     = {mar},
    number    = {3-4},
    pages     = {208--214},
    volume    = {59},
    doi       = {10.1007/BF01341426},
    publisher = {Springer Science and Business Media {LLC}}
}

@article{Gruneisen_ann.phys_16_530_1933,
    author={E. Gr{\"u}neisen},
    title = {{Die Abh{\"a}ngigkeit des elektrischen Widerstandes reiner Metalle von der Temperatur}},
    journal={Ann. Phys.},
    volume={408},
    pages = {530},
    year={1933},   
    doi = {10.1002/andp.19334080504},
    url = {https://doi.org/10.1002/andp.19334080504},
}

@article{lee_j.chem.phys_136_204120_2012,
    author = {Lee, Chee Kong and Moix, Jeremy and Cao, Jianshu},
    title = {Accuracy of second order perturbation theory in the polaron and variational polaron frames},
    journal = {J. Chem. Phys.},
    volume = {136},
    number = {20},
    pages = {204120},
    year = {2012},
    month = {05},
    issn = {0021-9606},
    doi = {10.1063/1.4722336},
    url = {https://doi.org/10.1063/1.4722336},
}

@article{stockburger_chem.phys.296.159.2004,
    title = {{Simulating spin-boson dynamics with stochastic Liouville–von Neumann equations}},
    journal = {Chem. Phys.},
    volume = {296},
    number = {2},
    pages = {159-169},
    year = {2004},
    issn = {0301-0104},
    doi = {https://doi.org/10.1016/j.chemphys.2003.09.014},
    url = {https://www.sciencedirect.com/science/article/pii/S0301010403004750},
    author = {Jürgen T. Stockburger}
}

@article{Donghwan_phys.rev.b_106_054311_2022,
    title = {Coherent charge carrier dynamics in the presence of thermal lattice vibrations},
    author = {Kim, Donghwan and Aydin, Alhun and Daza, Alvar and Avanaki, Kobra N. and Keski-Rahkonen, Joonas and Heller, Eric J.},
    journal = {Phys. Rev. B},
    volume = {106},
    issue = {5},
    pages = {054311},
    numpages = {25},
    year = {2022},
    month = {Aug},
    publisher = {American Physical Society},
    doi = {10.1103/PhysRevB.106.054311},
    url = {https://link.aps.org/doi/10.1103/PhysRevB.106.054311}
}

@article{HanburyBrown_proc.r.soc.a_242_200_1957,
    author = {R. Hanbury Brown and R. Q. Twiss},
    fjournal = {Proceedings of the Royal Society of London. Series A, Mathematical and Physical Sciences},
    journal = {Proc. R. Soc. A},
    number = {1230},
    pages = {300--324},
    publisher = {The Royal Society},
    title = {Interferometry of the Intensity Fluctuations in Light. I. Basic Theory: The Correlation between Photons in Coherent Beams of Radiation},
    volume = {242},
    year = {1957}
}

@article{Shockley_phys.rev_77_407_1950,
    title = {Energy Bands and Mobilities in Monatomic Semiconductors},
    author = {Shockley, W. and Bardeen, J.},
    journal = {Phys. Rev.},
    volume = {77},
    issue = {3},
    pages = {407--408},
    numpages = {0},
    year = {1950},
    month = {Feb},
    publisher = {American Physical Society},
    doi = {10.1103/PhysRev.77.407},
    url = {https://link.aps.org/doi/10.1103/PhysRev.77.407}
}

@article{Bardeen_phys.rev_80_72_1950,
    journal = {Phys. Rev.},
    author = {Bardeen, J. and Shockley, W.},
    volume = {80},
    issue = {1},
    pages = {72--80},
    numpages = {0},
    year = {1950},
    month = {Oct},
    title = {Deformation Potentials and Mobilities in Non-Polar Crystals},
    publisher = {American Physical Society},
    doi = {10.1103/PhysRev.80.72},
    url = {https://link.aps.org/doi/10.1103/PhysRev.80.72}
}

@article{aydin_pnas_122_e2426518122_2025,
    title={Polaron catastrophe within quantum acoustics},
    author={Aydin, Alhun and Keski-Rahkonen, Joonas and Graf, Anton M and Yuan, Shaobing and Ouyang, Xiao-Yu and M{\"u}stecapl{\i}o{\u{g}}lu, {\"O}zg{\"u}r E and Heller, Eric J},
    journal={Proc. Natl. Acad. Sci.},
    volume={122},
    number={23},
    pages={e2426518122},
    year={2025},
    publisher={National Academy of Sciences},
    doi = {10.1073/pnas.2426518122}
}

@article{Donghwan_phys.rev.B_107_224311_2023,
    title = {Low-energy tail of the spectral density for a particle interacting with a quantum phonon bath},
    author = {Kim, Donghwan and Halperin, Bertrand I.},
    journal = {Phys. Rev. B},
    volume = {107},
    issue = {22},
    pages = {224311},
    numpages = {18},
    year = {2023},
    month = {Jun},
    publisher = {American Physical Society},
    doi = {10.1103/PhysRevB.107.224311},
    url = {https://link.aps.org/doi/10.1103/PhysRevB.107.224311}
}

@article{aydin_pnas_121_e2404853121_2024,
    author = {Alhun Aydin  and Joonas Keski-Rahkonen  and Eric J. Heller },
    title = {{Quantum acoustics unravels Planckian resistivity}},
    journal = {Proc. Natl. Acad. Sci.},
    volume = {121},
    number = {28},
    pages = {e2404853121},
    year = {2024},
    doi = {10.1073/pnas.2404853121},
    URL = {https://www.pnas.org/doi/abs/10.1073/pnas.2404853121}
}

@article{KeskiRahkonen_phys.rev.lett_132_186303_2024,
    title = {{Quantum-acoustical Drude peak shift}},
    author = {Keski-Rahkonen, Joonas and Ouyang, Xiaoyu and Yuan, Shaobing and Graf, Anton M. and Aydin, Alhun and Heller, Eric J.},
    journal = {Phys. Rev. Lett.},
    volume = {132},
    issue = {18},
    pages = {186303},
    numpages = {8},
    year = {2024},
    month = {May},
    publisher = {American Physical Society},
    doi = {10.1103/PhysRevLett.132.186303},
    url = {https://link.aps.org/doi/10.1103/PhysRevLett.132.186303}
}

@article{Stockburger_vehm.phys_268_249_2001,
    title = {{Non-Markovian quantum state diffusion}},
    fjournal = {Chemical Physics},
    journal = {Chem. Phys.},
    volume = {268},
    number = {1},
    pages = {249-256},
    year = {2001},
    issn = {0301-0104},
    doi = {https://doi.org/10.1016/S0301-0104(01)00307-X},
    url = {https://www.sciencedirect.com/science/article/pii/S030101040100307X},
    author = {J{\"u}rgen T. Stockburger and Hermann Grabert}
}

@article{Frochlich_adv.phys_3_325_1954,
    author = { H. Fr{\"o}hlich },
    title = {Electrons in lattice fields},
    journal = {Adv. Phys.},
    fjournal = {Advances in Physics},
    volume = {3},
    number = {11},
    pages = {325-361},
    year  = {1954},
    publisher = {Taylor & Francis},
    doi = {10.1080/00018735400101213},
    URL = {https://doi.org/10.1080/00018735400101213}
}

@article{Frochlich_proc.r.soc.a_160_230_1937,
    author = {Fr{\"o}hlich, Herbert  and Mott, Nevill Francis},
    title = {Theory of electrical breakdown in ionic crystals},
    fjournal = {Proceedings of the Royal Society of London. Series A - Mathematical and Physical Sciences},
    journal = {Proc. R. Soc. A},
    volume = {160},
    number = {901},
    pages = {230-241},
    year = {1937},
    doi = {10.1098/rspa.1937.0106},
    URL = {https://royalsocietypublishing.org/doi/abs/10.1098/rspa.1937.0106}
}

@article{heller_J.Phys.Chem.A_123_4379_2019,
    title={Schr{\"o}dinger Correspondence Applied to Crystals},
    author={Heller, Eric J and Kim, Donghwan},
    fjournal={The Journal of Physical Chemistry A},
    journal ={J. Phys. Chem. A},
    volume={123},
    number={20},
    pages={4379--4388},
    year={2019},
    publisher={ACS Publications},
    doi={10.1021/acs.jpca.8b11746}
}

@article{Feynman_ann.phys_24_118_1963,
    title = {The theory of a general quantum system interacting with a linear dissipative system},
    journal = {Ann. Phys.},
    volume = {24},
    pages = {118-173},
    year = {1963},
    issn = {0003-4916},
    doi = {https://doi.org/10.1016/0003-4916(63)90068-X},
    author = {R.P Feynman and F.L Vernon}
}

@article{Hubbard_phys.rev.lett_3_77_1959,
    title = {Calculation of Partition Functions},
    author = {Hubbard, J.},
    journal = {Phys. Rev. Lett.},
    volume = {3},
    issue = {2},
    pages = {77--78},
    numpages = {0},
    year = {1959},
    month = {Jul},
    publisher = {American Physical Society},
    doi = {10.1103/PhysRevLett.3.77},
    url = {https://link.aps.org/doi/10.1103/PhysRevLett.3.77}
}

@inproceedings{stratonovich_book,
    title={On a method of calculating quantum distribution functions},
    author={Stratonovich, RL},
    booktitle={Soviet Physics Doklady}, 
    publisher = {American Institute of Physics (AIP)},
    volume={2},
    pages={416},
    year={1957},
    address={New York, NY, USA} 
}

@article{zimmermann_entropy_26_552_2024,
    title={{Rise and fall of Anderson localization by lattice vibrations: A time-dependent machine learning approach}},
    author={Zimmermann, Yoel and Keski-Rahkonen, Joonas and Graf, Anton M and Heller, Eric J},
    journal={Entropy},
    volume={26},
    number={7},
    pages={552},
    year={2024},
    doi = {10.3390/e26070552},
    publisher={MDPI}
}

@article{Herrero_j.condens.matter.phys_26_233201_2014,
    title={Path-integral simulation of solids},
    author={Herrero, Carlos P and Ramirez, Rafael},
    journal = {	J. Condens. Matter Phys.},
    fjournal={Journal of Physics: Condensed Matter},
    volume={26},
    number={23},
    pages={233201},
    year={2014},
    publisher={IOP Publishing},
    doi = {10.1088/0953-8984/26/23/233201}
}

@article{Li_phys.rev.B_104_195201_2021,
    title = {Deformation potential extraction and computationally efficient mobility calculations in silicon from first principles},
    author = {Li, Zhen and Graziosi, Patrizio and Neophytou, Neophytos},
    journal = {Phys. Rev. B},
    volume = {104},
    issue = {19},
    pages = {195201},
    numpages = {18},
    year = {2021},
    month = {Nov},
    publisher = {American Physical Society},
    doi = {10.1103/PhysRevB.104.195201},
    url = {https://link.aps.org/doi/10.1103/PhysRevB.104.195201}
}

@article{Resta_phys.rev.b_44_11035_1991,
    title = {Deformation-potential theorem in metals and in dielectrics},
    author = {Resta, R.},
    journal = {Phys. Rev. B},
    volume = {44},
    issue = {20},
    pages = {11035--11041},
    numpages = {0},
    year = {1991},
    month = {Nov},
    publisher = {American Physical Society},
    doi = {10.1103/PhysRevB.44.11035},
    url = {https://link.aps.org/doi/10.1103/PhysRevB.44.11035}
}

@article{Jin_npj_9_190_2023,
    title={High-throughput deformation potential and electrical transport calculations},
    author={Jin, Yeqing and Wang, Xiangdong and Yao, Mingjia and Qiu, Di and Singh, David J and Xi, Jinyang and Yang, Jiong and Xi, Lili},
    journal={npj Comput. Mater.},
    volume={9},
    number={1},
    pages={190},
    year={2023},
    publisher={Nature Publishing Group UK London},
    doi = {10.1038/s41524-023-01153-x}
}

@Article{feld2026phonon,
author={Feld, Leon G.
and Boehme, Simon C.
and Sabisch, Sebastian
and Frenkel, Nadav
and Yazdani, Nuri
and Morad, Viktoriia
and Zhu, Chenglian
and Kim, Taehee
and Canossa, Stefano
and Svyrydenko, Mariia
and Tao, Rui
and Bodnarchuk, Maryna I.
and Lubin, Gur
and Kazes, Miri
and Wood, Vanessa
and Oron, Dan
and Rain{\`o}, Gabriele
and Kovalenko, Maksym V.},
title={Phonon-driven wavefunction localization enhances room-temperature single-photon purity in large hybrid lead halide perovskite quantum dots},
journal={Nature Communications},
year={2026},
month={Jan},
day={23},
issn={2041-1723},
doi={10.1038/s41467-026-68607-w},
}

@article{Feldmann_jacs_143_8647_2021,
    title={Charge carrier localization in doped perovskite nanocrystals enhances radiative recombination},
    author={Feldmann, Sascha and Gangishetty, Mahesh K and Bravi{\'c}, Ivona and Neumann, Timo and Peng, Bo and Winkler, Thomas and Friend, Richard H and Monserrat, Bartomeu and Congreve, Daniel N and Deschler, Felix},
    journal={J. Am. Chem. Soc.},
    volume={143},
    number={23},
    pages={8647--8653},
    year={2021},
    publisher={ACS Publications},
    doi = {10.1021/jacs.1c01567}
}

@article{Blach_nano.lett_22_7811_2022,
    title={Superradiance and exciton delocalization in perovskite quantum dot superlattices},
    author={Blach, Daria D and Lumsargis, Victoria A and Clark, Daniel E and Chuang, Chern and Wang, Kang and Dou, Letian and Schaller, Richard D and Cao, Jianshu and Li, Christina W and Huang, Libai},
    journal = {Nano Lett.},
    fjournal={Nano Letters},
    volume={22},
    number={19},
    pages={7811--7818},
    year={2022},
    publisher={ACS Publications},
    doi = {10.1021/acs.nanolett.2c02427}
}

@article{Biswas_phys.rev.lett_125_076403_2020,
    title={{Spin-reorientation-induced band gap in $\textrm{Fe}_3\textrm{Sn}_2$: Optical signatures of Weyl nodes}},
    author={Biswas, A and Iakutkina, O and Wang, Q and Lei, HC and Dressel, M and Uykur, E},
    journal = {Phys. Rev. Lett.},
    fjournal={Physical Review Letters},
    volume={125},
    number={7},
    pages={076403},
    year={2020},
    publisher={APS},
    doi = {10.1103/PhysRevLett.125.076403}
}

@article{Pustogow_nat.comm_12_1571_2021,
    title={{Rise and fall of Landau’s quasiparticles while approaching the Mott transition}},
    author={Pustogow, Andrej and Saito, Yohei and L{\"o}hle, Anja and Sanz Alonso, Miriam and Kawamoto, Atsushi and Dobrosavljevi{\'c}, Vladimir and Dressel, Martin and Fratini, Simone},
    journal={Nat. Comm.},
    volume={12},
    number={1},
    pages={1571},
    year={2021},
    publisher={Nature Publishing Group UK London},
    doi = {10.1038/s41467-021-21741-z}
}

@article{Jaramillo_nat.phys_10_304_2014,
    title={Origins of bad-metal conductivity and the insulator--metal transition in the rare-earth nickelates},
    author={Jaramillo, R and Ha, Sieu D and Silevitch, DM and Ramanathan, Shriram},
    journal={Nat. Phys.},
    volume={10},
    number={4},
    pages={304--307},
    year={2014},
    publisher={Nature Publishing Group UK London},
    doi = {10.1038/nphys2907}
}

@article{Michon_nat.comm_14_3033_2023,
    title={{Reconciling scaling of the optical conductivity of cuprate superconductors with Planckian resistivity and specific heat}},
    author={Michon, Bastien and Berthod, Christophe and Rischau, Carl Willem and Ataei, Amirreza and Chen, Lu and Komiya, Seiki and Ono, Shimpei and Taillefer, Louis and van der Marel, Dirk and Georges, Antoine},
    journal={Nat. Comm.},
    volume={14},
    number={1},
    pages={3033},
    year={2023},
    publisher={Nature Publishing Group UK London},
    doi = {10.1038/s41467-023-38762-5}
}

@article{Varma_rev.mod.phys_92_031001_2020,
    author    = {Chandra M. Varma},
    title     = {Colloquium : Linear in temperature resistivity and associated mysteries including high temperature superconductivity},
    journal = {Rev. Mod. Phys.},
    year      = {2020},
    month     = {jul},
    number    = {3},
    pages     = {031001},
    volume    = {92},
    doi       = {10.1103/RevModPhys.92.031001},
    publisher = {American Physical Society ({APS})}
}

@article{Bruin_science_339_804_2013,
    author = {J. A. N. Bruin  and H. Sakai  and R. S. Perry  and A. P. Mackenzie },
    title = {Similarity of Scattering Rates in Metals Showing \textit{T}-Linear Resistivity},
    journal = {Science},
    volume = {339},
    number = {6121},
    pages = {804-807},
    year = {2013},
    doi = {10.1126/science.1227612},
    URL = {https://www.science.org/doi/abs/10.1126/science.1227612}
}

@article{Polshyn_nat.phys_15_1011_2019,
    author={Polshyn, Hryhoriy
    and Yankowitz, Matthew
    and Chen, Shaowen
    and Zhang, Yuxuan
    and Watanabe, K.
    and Taniguchi, T.
    and Dean, Cory R.
    and Young, Andrea F.},
    title={Large linear-in-temperature resistivity in twisted bilayer graphene},
    journal={Nat. Phys.},
    year={2019},
    month={Oct},
    day={01},
    volume={15},
    number={10},
    pages={1011-1016},
    issn={1745-2481},
    doi={10.1038/s41567-019-0596-3},
    url={https://doi.org/10.1038/s41567-019-0596-3}
}

@article{Vasiliu_phys.rev.lett_83_4393_1999,
    title={Charge melting and polaron collapse in $\mathrm{La}_{1.2}\mathrm{Sr}_{1.8}\mathrm{Mn}_2\mathrm{O}_7$},
    author={Vasiliu-Doloc, L and Rosenkranz, S and Osborn, R and Sinha, SK and Lynn, JW and Mesot, J and Seeck, OH and Preosti, G and Fedro, AJ and Mitchell, JF},
    journal={Phys. Rev. Lett.},
    volume={83},
    number={21},
    pages={4393},
    year={1999},
    publisher={APS},
    doi = {10.1103/PhysRevLett.83.4393}
}

@article{Guzelturk_nat.mater_20_618_2021,
    title={Visualization of dynamic polaronic strain fields in hybrid lead halide perovskites},
    author={Guzelturk, Burak and Winkler, Thomas and Van de Goor, Tim WJ and Smith, Matthew D and Bourelle, Sean A and Feldmann, Sascha and Trigo, Mariano and Teitelbaum, Samuel W and Steinr{\"u}ck, Hans-Georg and de la Pena, Gilberto A and others},
    journal={Nat. Mater.},
    volume={20},
    number={5},
    pages={618--623},
    year={2021},
    publisher={Nature Publishing Group UK London},
    doi = {10.1038/s41563-020-00865-5}
}

@article{Hwang_phys.rev.b_77_115449_2008,
    title = {Acoustic phonon scattering limited carrier mobility in two-dimensional extrinsic graphene},
    author = {Hwang, E. H. and Das Sarma, S.},
    journal = {Phys. Rev. B},
    volume = {77},
    issue = {11},
    pages = {115449},
    numpages = {6},
    year = {2008},
    month = {Mar},
    publisher = {American Physical Society},
    doi = {10.1103/PhysRevB.77.115449},
    url = {https://link.aps.org/doi/10.1103/PhysRevB.77.115449}
}

@article{Kim_phys.rev.lett_105_256805_2010,
    title = {Controlling Electron-Phonon Interactions in Graphene at Ultrahigh Carrier Densities},
    author = {Efetov, Dmitri K. and Kim, Philip},
    journal = {Phys. Rev. Lett.},
    volume = {105},
    issue = {25},
    pages = {256805},
    numpages = {4},
    year = {2010},
    month = {Dec},
    publisher = {American Physical Society},
    doi = {10.1103/PhysRevLett.105.256805},
    url = {https://link.aps.org/doi/10.1103/PhysRevLett.105.256805}
}

@article{Ciuchi_phys.rev.B_83_081202_2011,
    title = {Transient localization in crystalline organic semiconductors},
    author = {Ciuchi, S. and Fratini, S. and Mayou, D.},
    journal = {Phys. Rev. B},
    volume = {83},
    issue = {8},
    pages = {081202},
    numpages = {4},
    year = {2011},
    month = {Feb},
    publisher = {American Physical Society},
    doi = {10.1103/PhysRevB.83.081202},
    url = {https://link.aps.org/doi/10.1103/PhysRevB.83.081202}
}

@article{Ghost_of_Anderson_Arxiv,
    title={Planckian Diffusion: The Ghost of {A}nderson Localization},
    author={Zhang, Yubo and Graf, Anton M and Aydin, Alhun and Keski-Rahkonen, Joonas and Heller, Eric J},
    journal={arXiv preprint arXiv:2411.18768},
    url = {https://arxiv.org/pdf/2411.18768},
    year={2024}
}

@article{stockburger_phys.rev.lett_88_170407_2002,
    title = {Exact $\mathit{c}$-Number Representation of Non-Markovian Quantum Dissipation},
    author = {Stockburger, J\"urgen T. and Grabert, Hermann},
    journal = {Phys. Rev. Lett.},
    volume = {88},
    issue = {17},
    pages = {170407},
    numpages = {4},
    year = {2002},
    month = {Apr},
    publisher = {American Physical Society},
    doi = {10.1103/PhysRevLett.88.170407},
    url = {https://link.aps.org/doi/10.1103/PhysRevLett.88.170407}
}

@article{baihua_j.chem.theory.comput_21_3775_2025,
    author = {Wu, Baihua and Li, Bingqi and He, Xin and Cheng, Xiangsong and Ren, Jiajun and Liu, Jian},
    title = {Nonadiabatic Field: A Conceptually Novel Approach for Nonadiabatic Quantum Molecular Dynamics},
    journal = {J. Chem. Theory Comput.},
    fjournal = {Journal of Chemical Theory and Computation},
    volume = {21},
    number = {8},
    pages = {3775-3813},
    year = {2025},
    doi = {10.1021/acs.jctc.5c00181}
}

@article{zhang_j.chem.theory.comput_216352_2025,
    author = {Zhang, Jiaji and Liu, Jian and Chen, Lipeng},
    title = {Twin-Space Representation of Classical Mapping Model in the Constraint Phase Space Representation: Numerically Exact Approach to Open Quantum Systems},
    journal = {J. Chem. Theory Comput.},
    fjournal = {Journal of Chemical Theory and Computation},
    volume = {21},
    number = {13},
    pages = {6352-6366},
    year = {2025},
    doi = {10.1021/acs.jctc.5c00224}
}

@article{superwire1,
    author = {Alvar Daza  and Eric J. Heller  and Anton M. Graf  and Esa Räsänen },
    title = {Propagation of waves in high Brillouin zones: Chaotic branched flow and stable superwires},
    journal = {Proceedings of the National Academy of Sciences},
    volume = {118},
    number = {40},
    pages = {e2110285118},
    year = {2021},
    doi = {10.1073/pnas.2110285118},
    URL = {https://www.pnas.org/doi/abs/10.1073/pnas.2110285118}
}

@article{superwire2,
    title={Chaos-Assisted Dynamical Tunneling in Flat Band Superwires},
    author={Graf, Anton M and Lin, Ke and Kim, MyeongSeo and Keski-Rahkonen, Joonas and Daza, Alvar and Heller, Eric J},
    journal={Entropy},
    volume={26},
    number={6},
    pages={492},
    year={2024},
    publisher={MDPI}
}

@book{ziman2001electrons,
  title={Electrons and Phonons: The Theory of Transport Phenomena in Solids},
  author={Ziman, J.M.},
  isbn={9780198507796},
  lccn={2001268325},
  series={International series of monographs on physics},
  url={https://books.google.com/books?id=UtEy63pjngsC},
  year={2001},
  publisher={OUP Oxford},
  address={Oxford, UK}
}

@article{keski2024antiscarring,
  title={Antiscarring in Chaotic Quantum Wells},
  author={Keski-Rahkonen, J and Graf, AM and Heller, EJ},
  journal={arXiv preprint arXiv:2403.18081},
  year={2024},
  eprint={2403.18081}
}

@article{keski2019quantum,
  title={Quantum {L}issajous scars},
  author={Keski-Rahkonen, J and Ruhanen, A and Heller, EJ and R{\"a}s{\"a}nen, E},
  journal={Phys. Rev. Lett.},
  volume={123},
  number={21},
  pages={214101},
  year={2019},
  publisher={APS},
  doi = {10.1103/PhysRevLett.123.214101},
}

@book{heller2018semiclassical,
    title={The semiclassical way to dynamics and spectroscopy},
    author={Heller, Eric J},
    year={2018},
    publisher={Princeton University Press},
    doi={10.23943/9781400890293},
    address={Princeton, NJ, USA} 
}

@book{tannor2007introduction,
    title={Introduction to Quantum Mechanics},
    author={Tannor, D.J.},
    isbn={9781891389238},
    lccn={2006043933},
    url={https://books.google.com/books?id=WbQOzQEACAAJ},
    year={2007},
    publisher={University Science Books},
    address={Sausalito, CA, USA} 
}

@article{keskirahkonen_PhysRevB.96.094204.2017,
  title = {Controllable quantum scars in semiconductor quantum dots},
  author = {Keski-Rahkonen, J. and Luukko, P. J. J. and Kaplan, L. and Heller, E. J. and R\"as\"anen, E.},
  journal = {Phys. Rev. B},
  volume = {96},
  issue = {9},
  pages = {094204},
  numpages = {4},
  year = {2017},
  month = {Sep},
  publisher = {American Physical Society},
  doi = {10.1103/PhysRevB.96.094204},
  url = {https://link.aps.org/doi/10.1103/PhysRevB.96.094204}
}

@article{heller_nano.lett_5_1285_2005,
    title={Thermal averages in a quantum point contact with a single coherent wave packet},
    author={Heller, EJ and Aidala, KE and LeRoy, BJ and Bleszynski, AC and Kalben, A and Westervelt, RM and Maranowski, KD and Gossard, AC},
    journal={Nano Lett.},
    volume={5},
    number={7},
    pages={1285--1292},
    year={2005},
    publisher={ACS Publications},
    url = {https://doi.org/10.1021/nl0504585},
    doi = {10.1021/nl0504585}
}

@article{StockburgerColloquium,
  title = {Colloquium: Simulating non-Markovian dynamics in open quantum systems},
  author = {Xu, Meng and Vadimov, Vasilii and Stockburger, J. T. and Ankerhold, J.},
  journal = {Rev. Mod. Phys.},
  volume = {98},
  issue = {2},
  pages = {021002},
  numpages = {26},
  year = {2026},
  month = {May},
  publisher = {American Physical Society},
  doi = {10.1103/w3nw-hbjc},
  url = {https://link.aps.org/doi/10.1103/w3nw-hbjc}
}

@article{Egger_DynamicalSignProblem,
  title = {Path-integral Monte Carlo simulations without the sign problem: Multilevel blocking approach for effective actions},
  author = {Egger, R. and M\"uhlbacher, L. and Mak, C. H.},
  journal = {Phys. Rev. E},
  volume = {61},
  issue = {5},
  pages = {5961--5966},
  numpages = {0},
  year = {2000},
  month = {May},
  publisher = {American Physical Society},
  doi = {10.1103/PhysRevE.61.5961},
  url = {https://link.aps.org/doi/10.1103/PhysRevE.61.5961}
}

@article{BarkerPathIntegralMC,
    author = {Barker, J. A.},
    title = {A quantum‐statistical Monte Carlo method; path integrals with boundary conditions},
    journal = {The Journal of Chemical Physics},
    volume = {70},
    number = {6},
    pages = {2914-2918},
    year = {1979},
    month = {03},
    issn = {0021-9606},
    doi = {10.1063/1.437829},
    url = {https://doi.org/10.1063/1.437829}
}

@article{Althorpe_Annu.Rev.Phys.Chem_75_397_2024,
    title={Path integral simulations of condensed-phase vibrational spectroscopy},
    author={Althorpe, Stuart C},
    journal = {Annu. Rev. Phys. Chem.},
    fjournal={Annual Review of Physical Chemistry},
    volume={75},
    number={1},
    pages={397--420},
    year={2024},
    publisher={Annual Reviews},
    doi = {10.1146/annurev-physchem-090722-124705},
    url = {https://doi.org/10.1146/annurev-physchem-090722-124705}
}

@article{Markland_Nat.Rev.Chem_2_0109_2018,
    title={Nuclear quantum effects enter the mainstream},
    author={Markland, Thomas E and Ceriotti, Michele},
    journal={Nat. Rev. Chem.},
    volume={2},
    number={3},
    pages={0109},
    year={2018},
    publisher={Nature Publishing Group UK London},
    doi = {10.1038/s41570-017-0109},
    url = {https://doi.org/10.1038/s41570-017-0109}
}

@article{Ananth_Annu.Rev.Phys.Chem_73_299_2022,
    title={Path integrals for nonadiabatic dynamics: multistate ring polymer molecular dynamics},
    author={Ananth, Nandini},
    journal={Annu. Rev. Phys. Chem.},
    volume={73},
    pages={299--322},
    year={2022},
    publisher={Annual Reviews},
    doi = {10.1146/annurev-physchem-082620-021809},
    url = {https://doi.org/10.1146/annurev-physchem-082620-021809}
}

@article{Wong_Commun.Comput.Phys_15_853_2014,
    title={{Review of Feynman’s path integral in quantum statistics: From the molecular Schr{\"o}dinger equation to Kleinert’s variational perturbation theory}},
    author={Wong, Kin-Yiu},
    journal={Commun. Comput. Phys.},
    volume={15},
    number={4},
    pages={853--894},
    year={2014},
    publisher={Cambridge University Press},
    doi = {10.4208/cicp.140313.070513s },
    url = {https://doi.org/10.4208/cicp.140313.070513s }
}

@article{Legget_Rev.Mod.Phys_59_1_1987,
    title = {Dynamics of the dissipative two-state system},
    author = {Leggett, A. J. and Chakravarty, S. and Dorsey, A. T. and Fisher, Matthew P. A. and Garg, Anupam and Zwerger, W.},
    journal = {Rev. Mod. Phys.},
    volume = {59},
    issue = {1},
    pages = {1--85},
    numpages = {0},
    year = {1987},
    month = {Jan},
    publisher = {American Physical Society},
    doi = {10.1103/RevModPhys.59.1},
    url = {https://link.aps.org/doi/10.1103/RevModPhys.59.1}
}

@article{Grabert_Phys.Rep_168_115_1988,
    title={Quantum Brownian motion: The functional integral approach},
    author={Grabert, Hermann and Schramm, Peter and Ingold, Gert-Ludwig},
    journal={Phys. Rep.},
    volume={168},
    number={3},
    pages={115--207},
    year={1988},
    publisher={Elsevier},
    doi = {10.1016/0370-1573(88)90023-3},
    url = {https://doi.org/10.1016/0370-1573(88)90023-3}
}

@article{Sieberer_Rep.Prog.Phys_79_096001_2016,
    title={Keldysh field theory for driven open quantum systems},
    author={Sieberer, Lukas M and Buchhold, Michael and Diehl, Sebastian},
    journal={Rep. Prog. Phys},
    volume={79},
    number={9},
    pages={096001},
    year={2016},
    publisher={IOP Publishing},
    doi = {10.1088/0034-4885/79/9/096001},
    url = {https://doi.org/10.1088/0034-4885/79/9/096001}
}

@article{Dornheim_Phys.Rep_744_1_2018,
    title={The uniform electron gas at warm dense matter conditions},
    author={Dornheim, Tobias and Groth, Simon and Bonitz, Michael},
    journal={Phys. Rep.},
    volume={744},
    pages={1--86},
    year={2018},
    publisher={Elsevier},
    doi = {10.1016/j.physrep.2018.04.001},
    url = {https://doi.org/10.1016/j.physrep.2018.04.001}
}

@article{Ceperley_Rev.Mod.Phys_67_279_1995,
    title = {Path integrals in the theory of condensed helium},
    author = {Ceperley, D. M.},
    journal = {Rev. Mod. Phys.},
    volume = {67},
    issue = {2},
    pages = {279--355},
    numpages = {0},
    year = {1995},
    month = {Apr},
    publisher = {American Physical Society},
    doi = {10.1103/RevModPhys.67.279},
    url = {https://link.aps.org/doi/10.1103/RevModPhys.67.279}
}

@article{Ceperley_J.Stat.Phys_63_1237_1991,
    title={Fermion nodes},
    author={Ceperley, David M},
    journal={J. Stat. Phys.},
    volume={63},
    number={5},
    pages={1237--1267},
    year={1991},
    publisher={Springer},
    doi = {10.1007/BF01030009},
    url = {https://doi.org/10.1007/BF01030009}
}

@article{Feynman_Int.J.Theor.Phys_21_467_1982,
    author  = {Feynman, Richard P.},
    title   = {Simulating physics with computers},
    journal = {Int. J. Theor. Phys.},
    fjournal = {International Journal of Theoretical Physics},
    year    = {1982},
    volume  = {21},
    pages   = {467--488},
    doi     = {10.1007/BF02650179},
    url     = {https://doi.org/10.1007/BF02650179}
}

@article{Fosdick_Phys.Rev_143_58_1966,
    title = {Path-Integral Calculation of the Two-Particle Slater Sum for ${\mathrm{He}}^{4}$},
    author = {Fosdick, Lloyd D. and Jordan, Harry F.},
    journal = {Phys. Rev.},
    volume = {143},
    issue = {1},
    pages = {58--66},
    numpages = {0},
    year = {1966},
    month = {Mar},
    publisher = {American Physical Society},
    doi = {10.1103/PhysRev.143.58},
    url = {https://link.aps.org/doi/10.1103/PhysRev.143.58}
}

@article{Zhai_J.Phys.Chem.Lett_17_4274_2026,
    title={Geometric Phase Effect in Thermodynamic Properties and in the Imaginary-Time Multi-Electronic-State Path Integral Formulation},
    author={Zhai, Yu and Shang, Youhao and Liu, Jian},
    journal = {J. Phys. Chem. Lett.},
    fjournal={The Journal of Physical Chemistry Letters},
    volume={17},
    number={15},
    pages={4274--4291},
    year={2026},
    publisher={ACS Publications},
    url = {https://doi.org/10.1021/acs.jpclett.6c00429},
    doi = {10.1021/acs.jpclett.6c00429}
}

@article{Liu_J.Chem.Phys_148_102319_2018,
    title={Path integral molecular dynamics for exact quantum statistics of multi-electronic-state systems},
    author={Liu, Xinzijian and Liu, Jian},
    journal={J. Chem. Phys.},
    volume={148},
    number={10},
    pages = {102319},
    year={2018},
    publisher={AIP Publishing},
    url = {https://doi.org/10.1063/1.5005059},
    doi = {10.1063/1.5005059}
}

@article{Zhang_J.Chem.Phys_034109_2017,
    title={A unified thermostat scheme for efficient configurational sampling for classical/quantum canonical ensembles via molecular dynamics},
    author={Zhang, Zhijun and Liu, Xinzijian and Chen, Zifei and Zheng, Haifeng and Yan, Kangyu and Liu, Jian},
    journal={J. Chem. Phys.},
    volume={147},
    number={3},
    pages = {034109},
    year={2017},
    publisher={AIP Publishing},
    url = {https://doi.org/10.1063/1.4991621},
    doi = {10.1063/1.4991621}
}

@article{Liu_J.Chem.Phys_145_024103_2016,
    title={{A simple and accurate algorithm for path integral molecular dynamics with the Langevin thermostat}},
    author={Liu, Jian and Li, Dezhang and Liu, Xinzijian},
    journal={J. Chem. Phys.},
    volume={145},
    number={2},
    pages = {024103},
    year={2016},
    publisher={AIP Publishing},
    url ={https://doi.org/10.1063/1.4954990},
    doi = {10.1063/1.4954990}
}

@article{Rigol_Phys.Rev.Lett.108.110601_2012,
    title = {Alternatives to Eigenstate Thermalization},
    author = {Rigol, Marcos and Srednicki, Mark},
    journal = {Phys. Rev. Lett.},
    volume = {108},
    issue = {11},
    pages = {110601},
    numpages = {5},
    year = {2012},
    month = {Mar},
    publisher = {American Physical Society},
    doi = {10.1103/PhysRevLett.108.110601},
    url = {https://link.aps.org/doi/10.1103/PhysRevLett.108.110601}
}

@article{Srednicki_J.Phys.A_32_1163_1999,
    title={The approach to thermal equilibrium in quantized chaotic systems},
    author={Srednicki, Mark},
    journal={J. Phys. A},
    fjournal={Journal of Physics A: Mathematical and General},
    volume={32},
    number={7},
    pages={1163--1175},
    year={1999},
    url = {https://doi.org/10.1088/0305-4470/32/7/007},
    doi = {10.1088/0305-4470/32/7/007}
}

@article{Rigol_Nature_452_854_2008,
    title={Thermalization and its mechanism for generic isolated quantum systems},
    author={Rigol, Marcos and Dunjko, Vanja and Olshanii, Maxim},
    journal={Nature},
    volume={452},
    number={7189},
    pages={854--858},
    year={2008},
    publisher={Nature Publishing Group UK London},
    url = {https://doi.org/10.1038/nature06838},
    doi = {10.1038/nature06838}
}

@article{Srednicki_phys.rev.a_50_888_1994,
    title={Chaos and quantum thermalization},
    author={Srednicki, Mark},
    journal={Phys. Rev. E},
    volume={50},
    number={2},
    pages={888},
    year={1994},
    publisher={APS},
    url = {https://doi.org/10.1103/PhysRevE.50.888},
    doi ={10.1103/PhysRevE.50.888}
}

@article{Deutsch_phys.rev.a_43_2046_1991,
    title={Quantum statistical mechanics in a closed system},
    author={Deutsch, Josh M},
    journal={Phys. Rev. A},
    volume={43},
    number={4},
    pages={2046},
    year={1991},
    publisher={APS},
    url = {https://doi.org/10.1103/PhysRevA.43.2046},
    doi ={10.1103/PhysRevA.43.2046}
}

@article{Universal_quantum_birthmark_paper,
    title={Universal Quantum Birthmark: Ghost of the quantum past},
    author={Xiaoya, I. and Graf, A.M. and Heller, E. J. and Keski-Rahkonen, J.},
    journal={arXiv preprint arXiv:2602.00891},
    year={2026},
    url = {https://doi.org/10.48550/arXiv.2602.00891}
}

@article{Quantum_birthmark_paper,
    title={},
    author={Graf, A.M. and Atwood, S. and Xiao, M. and Ketzmerick, R. and Heller, E. J. and Keski-Rahkonen, J.},
    journal={arXiv preprint arXiv:2412.02982},
    year={2026},
    url = {https://doi.org/10.48550/arXiv.2412.02982}
}

@article{Zurek_Rev.Mod.Phys_75_715_2003,
    title = {Decoherence, einselection, and the quantum origins of the classical},
    author = {Zurek, Wojciech Hubert},
    journal = {Rev. Mod. Phys.},
    volume = {75},
    issue = {3},
    pages = {715--775},
    numpages = {0},
    year = {2003},
    month = {May},
    publisher = {American Physical Society},
    doi = {10.1103/RevModPhys.75.715},
    url = {https://link.aps.org/doi/10.1103/RevModPhys.75.715}
}

@article{Zurek_Nat.Phys_5_181_2009,
    title={{Quantum Darwinism}},
    author={Zurek, Wojciech Hubert},
    journal={Nat. Phys.},
    volume={5},
    number={3},
    pages={181--188},
    year={2009},
    publisher={Nature Publishing Group UK London},
    doi = {10.1038/nphys1202},
    url = {https://doi.org/10.1038/nphys1202}
}

@book{Zurek_book,
    author    = {Wojciech H. Zurek},
    title     = {Decoherence and Quantum Darwinism: From Quantum Foundations to Classical Reality},
    publisher = {Cambridge University Press},
    year       = {2025},
    isbn       = {9781009552868},
    doi        = {10.1017/9781009552868},
    address={Cambridge, UK}
}

@article{Rozenbaum_Phys.Rev.B_100_035112_2019,
    title = {Universal level statistics of the out-of-time-ordered operator},
    author = {Rozenbaum, Efim B. and Ganeshan, Sriram and Galitski, Victor},
    journal = {Phys. Rev. B},
    volume = {100},
    issue = {3},
    pages = {035112},
    numpages = {7},
    year = {2019},
    month = {Jul},
    publisher = {American Physical Society},
    doi = {10.1103/PhysRevB.100.035112},
    url = {https://link.aps.org/doi/10.1103/PhysRevB.100.035112}
}

@article{Jalabert_Phys.Rev.E_98_062218_2018,
    title = {Semiclassical theory of out-of-time-order correlators for low-dimensional classically chaotic systems},
    author = {Jalabert, Rodolfo A. and Garc\'{\i}a-Mata, Ignacio and Wisniacki, Diego A.},
    journal = {Phys. Rev. E},
    volume = {98},
    issue = {6},
    pages = {062218},
    numpages = {12},
    year = {2018},
    month = {Dec},
    publisher = {American Physical Society},
    doi = {10.1103/PhysRevE.98.062218},
    url = {https://link.aps.org/doi/10.1103/PhysRevE.98.062218}
}

@article{Hashimoto_J.High.Energy.Phys_2017_1_2017,
    title={Out-of-time-order correlators in quantum mechanics},
    author={Hashimoto, Koji and Murata, Keiju and Yoshii, Ryosuke},
    journal = {J. High Energy Phys.},
    ffournal={Journal of High Energy Physics},
    volume={2017},
    number={10},
    pages={1--31},
    year={2017},
    publisher={Springer},
    doi = {10.1007/JHEP10(2017)138},
    url = {https://doi.org/10.1007/JHEP10(2017)138}
}

@article{Jin_J.Chem.Phys_128_234703_2008,
    title={Exact dynamics of dissipative electronic systems and quantum transport: Hierarchical equations of motion approach},
    author={Jin, Jinshuang and Zheng, Xiao and Yan, YiJing},
    journal={J. Chem. Phys.},
    volume={128},
    number={23},
    pages = {234703},
    year={2008},
    publisher={AIP Publishing},
    doi = {10.1063/1.2938087},
    url = {https://doi.org/10.1063/1.2938087}
}

@article{Tanimura_J.Chem.Phys_153_020901,
    title={{Numerically ``exact'' approach to open quantum dynamics: The hierarchical equations of motion (HEOM)}},
    author={Tanimura, Yoshitaka},
    journal={J. Chem. Phys.},
    volume={153},
    number={2},
    pages = {020901},
    year={2020},
    publisher={AIP Publishing},
    doi = {10.1063/5.0011599},
    url = {https://doi.org/10.1063/5.0011599}
}

@article{Tanimura_J.Phys.Soc.Jpn_75_082001_2006,
    title={{Stochastic Liouville, Langevin, Fokker--Planck, and master equation approaches to quantum dissipative systems}},
    author={Tanimura, Yoshitaka},
    journal={J. Phys. Soc. Jpn},
    volume={75},
    number={8},
    pages={082001},
    year={2006},
    publisher={The Physical Society of Japan},
    url = {https://doi.org/10.1143/JPSJ.75.082001},
    doi = {10.1143/JPSJ.75.082001}
}

@article{Yan_J.Chem.Phys_140_054105_2014,
    title={Theory of open quantum systems with bath of electrons and phonons and spins: Many-dissipaton density matrixes approach},
    author={Yan, YiJing},
    journal={J. Chem. Phys.},
    volume={140},
    number={5},
    pages = {054105},
    year={2014},
    publisher={AIP Publishing},
    doi = {10.1063/1.4863379},
    url = {https://doi.org/10.1063/1.4863379}
}

@article{McCaul_Phys.Rev.B_95_125124_2017,
    title = {Partition-free approach to open quantum systems in harmonic environments: An exact stochastic Liouville equation},
    author = {McCaul, G. M. G. and Lorenz, C. D. and Kantorovich, L.},
    journal = {Phys. Rev. B},
    volume = {95},
    issue = {12},
    pages = {125124},
    numpages = {14},
    year = {2017},
    month = {Mar},
    publisher = {American Physical Society},
    doi = {10.1103/PhysRevB.95.125124},
    url = {https://link.aps.org/doi/10.1103/PhysRevB.95.125124}
}

@article{Pleasance_Phys.Rev.Research_2_043058_2020,
    title = {Generalized theory of pseudomodes for exact descriptions of non-Markovian quantum processes},
    author = {Pleasance, Graeme and Garraway, Barry M. and Petruccione, Francesco},
    journal = {Phys. Rev. Res.},
    volume = {2},
    issue = {4},
    pages = {043058},
    numpages = {12},
    year = {2020},
    month = {Oct},
    publisher = {American Physical Society},
    doi = {10.1103/PhysRevResearch.2.043058},
    url = {https://link.aps.org/doi/10.1103/PhysRevResearch.2.043058}
}

@article{Tamascelli_Phys.Rev.Lett_123_090402_2019,
    title = {Efficient Simulation of Finite-Temperature Open Quantum Systems},
    author = {Tamascelli, D. and Smirne, A. and Lim, J. and Huelga, S. F. and Plenio, M. B.},
    journal = {Phys. Rev. Lett.},
    volume = {123},
    issue = {9},
    pages = {090402},
    numpages = {6},
    year = {2019},
    month = {Aug},
    publisher = {American Physical Society},
    doi = {10.1103/PhysRevLett.123.090402},
    url = {https://link.aps.org/doi/10.1103/PhysRevLett.123.090402}
}

@article{Garraway_Phys.Rev.A_55_2290_1997,
    title = {Nonperturbative decay of an atomic system in a cavity},
    author = {Garraway, B. M.},
    journal = {Phys. Rev. A},
    volume = {55},
    issue = {3},
    pages = {2290--2303},
    numpages = {0},
    year = {1997},
    month = {Mar},
    publisher = {American Physical Society},
    doi = {10.1103/PhysRevA.55.2290},
    url = {https://link.aps.org/doi/10.1103/PhysRevA.55.2290}
}

@article{Garraway_Phys.Rev.A.54.3592_1996,
    title = {Cavity modified quantum beats},
    author = {Garraway, B. M. and Knight, P. L.},
    journal = {Phys. Rev. A},
    volume = {54},
    issue = {4},
    pages = {3592--3602},
    numpages = {0},
    year = {1996},
    month = {Oct},
    publisher = {American Physical Society},
    doi = {10.1103/PhysRevA.54.3592},
    url = {https://link.aps.org/doi/10.1103/PhysRevA.54.3592}
}

@article{Imamoglu_Phys.Rev.A_50_3650_1994,
    title = {Stochastic wave-function approach to non-Markovian systems},
    author = {Imamog\ifmmode\bar\else\textasciimacron\fi{}lu, A.},
    journal = {Phys. Rev. A},
    volume = {50},
    issue = {5},
    pages = {3650--3653},
    numpages = {0},
    year = {1994},
    month = {Nov},
    publisher = {American Physical Society},
    doi = {10.1103/PhysRevA.50.3650},
    url = {https://link.aps.org/doi/10.1103/PhysRevA.50.3650}
}

@article{Lentrodt_PhysRevLett.130.263602_2023,
    title = {Certifying Multimode Light-Matter Interaction in Lossy Resonators},
    author = {Lentrodt, Dominik and Diekmann, Oliver and Keitel, Christoph H. and Rotter, Stefan and Evers, J{\"o}rg},
    journal = {Phys. Rev. Lett.},
    volume = {130},
    issue = {26},
    pages = {263602},
    numpages = {9},
    year = {2023},
    month = {Jun},
    publisher = {American Physical Society},
    doi = {10.1103/PhysRevLett.130.263602},
    url = {https://link.aps.org/doi/10.1103/PhysRevLett.130.263602}
}

@article{Luo_PRXQuantum_4_030316_2023,
    title = {Quantum-Classical Decomposition of Gaussian Quantum Environments: A Stochastic Pseudomode Model},
    author = {Luo, Si and Lambert, Neill and Liang, Pengfei and Cirio, Mauro},
    journal = {PRX Quantum},
    volume = {4},
    issue = {3},
    pages = {030316},
    numpages = {43},
    year = {2023},
    month = {Aug},
    publisher = {American Physical Society},
    doi = {10.1103/PRXQuantum.4.030316},
    url = {https://link.aps.org/doi/10.1103/PRXQuantum.4.030316}
}

@article{Medina_Phys.Rev.Lett_126_093601_2021,
    title = {Few-Mode Field Quantization of Arbitrary Electromagnetic Spectral Densities},
    author = {Medina, Ivan and Garc\'{\i}a-Vidal, Francisco J. and Fern\'andez-Dom\'{\i}nguez, Antonio I. and Feist, Johannes},
    journal = {Phys. Rev. Lett.},
    volume = {126},
    issue = {9},
    pages = {093601},
    numpages = {7},
    year = {2021},
    month = {Mar},
    publisher = {American Physical Society},
    doi = {10.1103/PhysRevLett.126.093601},
    url = {https://link.aps.org/doi/10.1103/PhysRevLett.126.093601}
}

@article{Wang_J.Chem.Phys_110_4828_1999,
    title={Semiclassical study of electronically nonadiabatic dynamics in the condensed-phase: Spin-boson problem with Debye spectral density},
    author={Wang, Haobin and Song, Xueyu and Chandler, David and Miller, William H},
    journal={J. Chem. Phys.},
    volume={110},
    number={10},
    pages={4828--4840},
    year={1999},
    publisher={American Institute of Physics},
    doi = {10.1063/1.478388},
    url = {https://doi.org/10.1063/1.478388}
}

@article{Shi_J.Chem.Phys_163_224103_2025,
    title={Hierarchical equations of motion solved with the multiconfigurational Ehrenfest ansatz},
    author={Shi, Zhecun and Zhou, Huiqiang and Huang, Lei and Xie, Rixin and Wang, Linjun},
    journal={J. Chem. Phys.},
    volume={163},
    number={22},
    pages = {224103},
    year={2025},
    publisher={AIP Publishing},
    doi = {10.1063/5.0301142},
    url = { https://doi.org/10.1063/5.0301142}
}

@article{Stock_J.Chem.Phys_103_1561_1995,
    title={A semiclassical self-consistent-field approach to dissipative dynamics: The spin--boson problem},
    author={Stock, Gerhard},
    journal={J. Chem. Phys.},
    volume={103},
    number={4},
    pages={1561--1573},
    year={1995},
    publisher={American Institute of Physics},
    doi = {10.1063/1.469778},
    url = {https://doi.org/10.1063/1.469778}
}

@article{Shi_J.Chem.Phys_120_10647_2004,
    title={A semiclassical generalized quantum master equation for an arbitrary system-bath coupling},
    author={Shi, Qiang and Geva, Eitan},
    journal={J. Chem. Phys.},
    volume={120},
    number={22},
    pages={10647--10658},
    year={2004},
    publisher={American Institute of Physics},
    doi = {10.1063/1.1738109},
    url = {https://doi.org/10.1063/1.1738109}
}

@article{Provazza_J.Chem.Phys_151_154114_2019,
    title={Multi-level description of the vibronic dynamics of open quantum systems},
    author={Provazza, Justin and Coker, David F},
    journal={J. Chem. Phys.},
    volume={151},
    number={15},
    pages = {154114},
    year={2019},
    publisher={AIP Publishing},
    doi = {10.1063/1.5120253},
    url = {https://doi.org/10.1063/1.5120253}
}

@article{Martin_J.Chem.Phys_126_164108_2007,
    title={Semiclassical initial value series solution of the spin boson problem},
    author={Martin-Fierro, Eva and Pollak, Eli},
    journal={J. Chem. Phys.},
    volume={126},
    number={16},
    pages = {164108},
    year={2007},
    publisher={AIP Publishing},
    doi = {10.1063/1.2714520},
    url = {https://doi.org/10.1063/1.2714520}
}

@article{Koch_Phys.Rev.Lett_100_230402_2008,
    title = {Non-Markovian Dissipative Semiclassical Dynamics},
    author = {Koch, Werner and Gro\ss{}mann, Frank and Stockburger, J{\"u}rgen T. and Ankerhold, Joachim},
    journal = {Phys. Rev. Lett.},
    volume = {100},
    issue = {23},
    pages = {230402},
    numpages = {4},
    year = {2008},
    month = {Jun},
    publisher = {American Physical Society},
    doi = {10.1103/PhysRevLett.100.230402},
    url = {https://link.aps.org/doi/10.1103/PhysRevLett.100.230402}
}

@article{Hartmann_J.Chem.Theory.Comput_13_5834_2017,
    title={Exact open quantum system dynamics using the hierarchy of pure states (HOPS)},
    author={Hartmann, Richard and Strunz, Walter T},
    journal={J. Chem. Theory Comput.},
    volume={13},
    number={12},
    pages={5834--5845},
    year={2017},
    publisher={ACS Publications},
    doi = {10.1021/acs.jctc.7b00751},
    url = {https://doi.org/10.1021/acs.jctc.7b00751}
}

@article{Ritschel_J.Chem.Phys_142_034115_2015,
    title={Non-Markovian Quantum State Diffusion for temperature-dependent linear spectra of light harvesting aggregates},
    author={Ritschel, Gerhard and Suess, Daniel and M{\"o}bius, Sebastian and Strunz, Walter T and Eisfeld, Alexander},
    journal={J. Chem. Phys.},
    volume={142},
    number={3},
    pages = {034115},
    year={2015},
    publisher={AIP Publishing},
    doi = {10.1063/1.4905327},
    url = {https://doi.org/10.1063/1.4905327}
}

@article{Suess_J.Stat.Phys_159_1408_2015,
    title={Hierarchical equations for open system dynamics in fermionic and bosonic environments},
    author={Suess, Daniel and Strunz, Walter T and Eisfeld, Alexander},
    journal={J. Stat. Phys.},
    volume={159},
    number={6},
    pages={1408--1423},
    year={2015},
    publisher={Springer},
    doi = {10.1007/s10955-015-1236-7},
    url = {https://doi.org/10.1007/s10955-015-1236-7}
}

@article{Suess_Phys.Rev.Lett_113_150403_2014,
    title = {Hierarchy of Stochastic Pure States for Open Quantum System Dynamics},
    author = {Suess, D. and Eisfeld, A. and Strunz, W. T.},
    journal = {Phys. Rev. Lett.},
    volume = {113},
    issue = {15},
    pages = {150403},
    numpages = {5},
    year = {2014},
    month = {Oct},
    publisher = {American Physical Society},
    doi = {10.1103/PhysRevLett.113.150403},
    url = {https://link.aps.org/doi/10.1103/PhysRevLett.113.150403}
}

@article{Diosi_Phys.Rev.A_58_1699_1998,
    title = {Non-Markovian quantum state diffusion},
    author = {Di{\'o}si, L. and Gisin, N. and Strunz, W. T.},
    journal = {Phys. Rev. A},
    volume = {58},
    issue = {3},
    pages = {1699--1712},
    numpages = {0},
    year = {1998},
    month = {Sep},
    publisher = {American Physical Society},
    doi = {10.1103/PhysRevA.58.1699},
    url = {https://link.aps.org/doi/10.1103/PhysRevA.58.1699}
}

@article{Strunz_Phys.Rev.Lett_82_1801_1999,
    title = {Open System Dynamics with Non-Markovian Quantum Trajectories},
    author = {Strunz, Walter T. and Di\'osi, Lajos and Gisin, Nicolas},
    journal = {Phys. Rev. Lett.},
    volume = {82},
    issue = {9},
    pages = {1801--1805},
    numpages = {0},
    year = {1999},
    month = {Mar},
    publisher = {American Physical Society},
    doi = {10.1103/PhysRevLett.82.1801},
    url = {https://link.aps.org/doi/10.1103/PhysRevLett.82.1801}
}

@article{Borrelli_New.J.Phys_22_123002_2020,
    title={Quantum dynamics of vibrational energy flow in oscillator chains driven by anharmonic interactions},
    author={Borrelli, Raffaele and Gelin, Maxim F},
    journal={New J. Phys.},
    volume={22},
    number={12},
    pages={123002},
    year={2020},
    publisher={IOP Publishing},
    doi = {10.1088/1367-2630/abc9ed},
    url = {https://iopscience.iop.org/article/10.1088/1367-2630/abc9ed}
}

@article{Gelin_Ann.Phys_529_1700200_2017,
    title={Thermal schr{\"o}dinger equation: efficient tool for simulation of many-body quantum dynamics at finite temperature},
    author={Gelin, Maxim F and Borrelli, Raffaele},
    journal={Ann. Phys.},
    volume={529},
    number={12},
    pages={1700200},
    year={2017},
    publisher={Wiley Online Library},
    doi = {10.1002/andp.201700200},
    url = {https://doi.org/10.1002/andp.201700200}
}

@article{Gelin_J.Chem.Phys_155_134102_2021,
    title={Efficient quantum dynamics simulations of complex molecular systems: A unified treatment of dynamic and static disorder},
    author={Gelin, Maxim F and Velardo, Amalia and Borrelli, Raffaele},
    journal={J. Chem. Phys.},
    volume={155},
    pages = {134102},
    number={13},
    year={2021},
    publisher={AIP Publishing},
    doi = {10.1063/5.0065896},
    url = {https://doi.org/10.1063/5.0065896}
}

@article{Yu_Phys.Rev.A_69_062107_2004,
    title = {Non-Markovian quantum trajectories versus master equations: Finite-temperature heat bath},
    author = {Yu, Ting},
    journal = {Phys. Rev. A},
    volume = {69},
    issue = {6},
    pages = {062107},
    numpages = {9},
    year = {2004},
    month = {Jun},
    publisher = {American Physical Society},
    doi = {10.1103/PhysRevA.69.062107},
    url = {https://link.aps.org/doi/10.1103/PhysRevA.69.062107}
}

@article{de.Vega_Phys.Rev.A_92_052116_2015,
    title = {Thermofield-based chain-mapping approach for open quantum systems},
    author = {de Vega, In{\'e}s and Ba{\~n}uls, Mari-Carmen},
    journal = {Phys. Rev. A},
    volume = {92},
    issue = {5},
    pages = {052116},
    numpages = {7},
    year = {2015},
    month = {Nov},
    publisher = {American Physical Society},
    doi = {10.1103/PhysRevA.92.052116},
    url = {https://link.aps.org/doi/10.1103/PhysRevA.92.052116}
}

@article{Breuer_Rev.Mod.Phys_88_021002_2016,
    title = {Colloquium: Non-Markovian dynamics in open quantum systems},
    author = {Breuer, Heinz-Peter and Laine, Elsi-Mari and Piilo, Jyrki and Vacchini, Bassano},
    journal = {Rev. Mod. Phys.},
    volume = {88},
    issue = {2},
    pages = {021002},
    numpages = {24},
    year = {2016},
    month = {Apr},
    publisher = {American Physical Society},
    doi = {10.1103/RevModPhys.88.021002},
    url = {https://link.aps.org/doi/10.1103/RevModPhys.88.021002}
}

@article{Gonzalez_Quantum_8_1454_2024,
    title={Tutorial: projector approach to master equations for open quantum systems},
    author={Gonzalez-Ballestero, Carlos},
    journal={Quantum},
    volume={8},
    pages={1454},
    year={2024},
    publisher={Verein zur F{\"o}rderung des Open Access Publizierens in den Quantenwissenschaften},
    url = {https://doi.org/10.22331/q-2024-08-29-1454},
    doi = {10.22331/q-2024-08-29-1454}
}

@article{Mulvihill_J.Phys.Chem.B_125_9834_2021,
    title={A road map to various pathways for calculating the memory kernel of the generalized quantum master equation},
    author={Mulvihill, Ellen and Geva, Eitan},
    journal={ J. Phys. Chem. B},
    volume={125},
    number={34},
    pages={9834--9852},
    year={2021},
    publisher={ACS Publications},
    doi = {10.1021/acs.jpcb.1c05719},
    url = {https://doi.org/10.1021/acs.jpcb.1c05719}
}


\end{document}